\documentclass[aps,pre,showpacs,noshowkeys,amsmath,amssymb,amsfonts,superscriptaddress,longbibliography,reprint]{revtex4-2}
\usepackage[T1]{fontenc}
\usepackage{newtxtext}

\usepackage{enumitem}

\usepackage{physics}
\usepackage{amsmath}
\usepackage{mathtools}
\usepackage{dsfont}

\usepackage{appendix}
\usepackage{chngcntr}

\usepackage{placeins}

\usepackage{color}
\usepackage[dvipsnames]{xcolor}

\definecolor{nqdcolor}{rgb}{0.5586, 0.0586, 0.4219}
\newcommand*{\nqdcolor}{\color{nqdcolor}}

\usepackage{hyperref}
\usepackage[nameinlink,capitalise]{cleveref}
\crefname{extfigure}{Extended Data Fig.}{Extended Data Figs.}
\Crefname{extfigure}{Extended Data Fig.}{Extended Data Figs.}
\crefname{extsubfigure}{Extended Data Fig.}{Extended Data Figs.}
\Crefname{extsubfigure}{Extended Data Fig.}{Extended Data Figs.}

\hypersetup{colorlinks,citecolor=nqdcolor,linkcolor=nqdcolor,urlcolor=nqdcolor}

\usepackage{makecell}
\usepackage{colortbl} 
\usepackage{tabularx}

\usepackage{graphicx}
\usepackage[all]{hypcap}

\usepackage{subcaption}
\DeclareCaptionLabelSeparator{bar}{~\rule[-0.4ex]{0.2ex}{1em}~}
\DeclareCaptionLabelFormat{subfor}{\textbf{#2}}
\usepackage{comment}

\definecolor{lhcolor}{rgb}{0.2,0.5,0.4}

\newcommand{\nocontentsline}[3]{}
\let\origcontentsline\addcontentsline
\newcommand\stoptoc{\let\addcontentsline\nocontentsline}
\newcommand\resumetoc{\let\addcontentsline\origcontentsline}

\def\({\left(}
\def\){\right)}                       
\def\[{\left[}
\def\]{\right]}

\newcommand{\beq}{\begin{equation}}
\newcommand{\eeq}{\end{equation}}
\newcommand{\ba}{\begin{align}}
\newcommand{\ea}{\end{align}}

\newcommand{\tens}[1]{\mathbf{#1}}

\renewcommand{\ni}{\hat{\vec n}_i}
\renewcommand{\Re}{\text{Re}\,}
\renewcommand{\Im}{\text{Im}\,}

\usepackage{bm}	
\renewcommand{\vec}[1]{\bm{#1}}

\begin{document}

\author{Lauritz Hahn}
\affiliation{Max Planck Institute for the Physics of Complex Systems, N\"{o}thnitzer Str.~38, 01187 Dresden, Germany}
\affiliation{Center for Systems Biology Dresden, Pfotenhauerstr. 108, 01307 Dresden, Germany}

\author{Xinyu Dou}
\affiliation{Department of Polymer Science and Engineering, University of Science and Technology of China (USTC),  Hefei, Anhui 230026, China}
\affiliation{Key Laboratory of Precision and Intelligent Chemistry, University of Science and Technology of China (USTC), Hefei, Anhui 230026, China}

\author{Ziqi Zhou}
\affiliation{Department of Polymer Science and Engineering, University of Science and Technology of China (USTC),  Hefei, Anhui 230026, China}
\affiliation{Key Laboratory of Precision and Intelligent Chemistry, University of Science and Technology of China (USTC), Hefei, Anhui 230026, China}

\author{Jie Zhang}
\affiliation{Department of Polymer Science and Engineering, University of Science and Technology of China (USTC),  Hefei, Anhui 230026, China}
\affiliation{Key Laboratory of Precision and Intelligent Chemistry, University of Science and Technology of China (USTC), Hefei, Anhui 230026, China}

\author{Marin Bukov}
\thanks{Equal contribution; \nqdcolor{mgbukov@pks.mpg.de}, \nqdcolor{ricard.alert@ub.edu}}
\affiliation{Max Planck Institute for the Physics of Complex Systems, N\"{o}thnitzer Str.~38, 01187 Dresden, Germany}

\author{Ricard Alert}
\thanks{Equal contribution; \nqdcolor{mgbukov@pks.mpg.de}, \nqdcolor{ricard.alert@ub.edu}}
\affiliation{Max Planck Institute for the Physics of Complex Systems, N\"{o}thnitzer Str.~38, 01187 Dresden, Germany}
\affiliation{Center for Systems Biology Dresden, Pfotenhauerstr. 108, 01307 Dresden, Germany}
\affiliation{Departament de F\'{i}sica de la Mat\`{e}ria Condensada, Universitat de Barcelona, Barcelona, Spain}
\affiliation{Universitat de Barcelona Institute of Complex Systems (UBICS), Barcelona, Spain}
\affiliation{Instituci\'{o} Catalana de Recerca i Estudis Avan\c{c}ats (ICREA), Barcelona, Spain}

\clearpage
\title{Polar Streams through Active Condensation in Flocks} 


\begin{abstract}
		Two of the most paradigmatic phenomena in active matter are flocking, where particles align to move collectively, and motility-induced phase separation, where self-propulsion causes particles to condense even without attractive interactions. Here, using experiments on self-propelled Janus colloid dimers, we show that flocks can phase separate to form polar streams. These structures are elongated dense clusters of particles that flock along the stream's long axis, in contrast to the well-known transverse bands that form close to the flocking transition. We use particle-based simulations and hydrodynamic theory obtained through coarse-graining to identify the underlying mechanism, which we call active nematic condensation: Denser regions of the flock develop stronger polar and hence nematic order, which leads to further particle fluxes into the dense region. These results reveal a generic mechanism whereby flocks can condense into streams. Our findings uncover that flocks can be shaped by nematic order, which is usually not accounted for in the standard Toner-Tu equations. More broadly, our work shows that collective motion and phase separation can not only coexist but also cooperate in active matter.
\end{abstract}

\maketitle
	
\stoptoc
\section{Introduction}

Active matter exhibits striking non-equilibrium collective phenomena such as flocking \cite{vicsek2012,marchetti2013} and motility-induced phase separation (MIPS) \cite{cates2015}, which both arise due to the interplay of self-propulsion and interactions between active particles. Flocking --- the self-organized collective motion of active particles --- appears in many systems across scales, ranging from bird swarms \cite{cavagna2014} and cell layers \cite{szabo2006,malinverno2017,brandstatter2023,shen2025b,schwabach2026} to synthetic active systems like Quincke rollers \cite{bricard2013} and carpets of cytoskeletal filaments \cite{schaller2010}. Flocking is typically caused by polar alignment interactions between particles, as in the Vicsek model \cite{vicsek1995}, but can also be due to other effects such as turn-away torques \cite{das2024} and polarity-velocity self-alignment \cite{szabo2006, weber2013, lam2015a, baconnier2025}. Respectively, active phase separation can be caused by a decrease in motility of active agents in dense regions due to quorum sensing \cite{tailleur2008} or repulsive interactions \cite{fily2012}, but also due to other effects such as turn-towards torques \cite{zhang2021} or active nematic flows \cite{bhattacharyya2023}.

While these phenomena are often studied in isolation by means of simple models, in experiments they were found to compete with each other \cite{geyer2019,vanderlinden2019}. On the one hand, MIPS can cause flocks to jam and lose polar order \cite{geyer2019}. On the other hand, polar alignment was shown to interrupt motility-induced phase separation \cite{vanderlinden2019}. Thus, it remains unclear to what extent flocking and MIPS can coexist \cite{luo2025}.

Here, we show that flocks can undergo active phase separation to form polar streams, i.e., elongated dense clusters in which particles collectively move along the long axis of the cluster. We first observe polar streams in experiments with Janus colloid dimers. To explain this behavior, we develop and simulate a microscopic model of active particles that captures the emergence of streams through a flocking transition followed by phase separation into dense streams within a dilute flocking phase. To understand these transitions, we coarse-grain our model to derive a hydrodynamic theory, which explains the observations and allows us to identify the mechanism of stream formation. We show that streams are the result of phase separation caused by extensile nematic fluxes that condense the flock perpendicularly to its direction of motion --- a mechanism that we call active nematic condensation. Our results uncover physical mechanisms that shape flocks: we reveal how flocking and phase separation can take place sequentially to cooperatively organize a flock into dense and dilute phases.

\section{Streams of metal-dielectric Janus colloids}

\begin{figure*}[t!]
	\centering
	\includegraphics[width=1.0\textwidth]{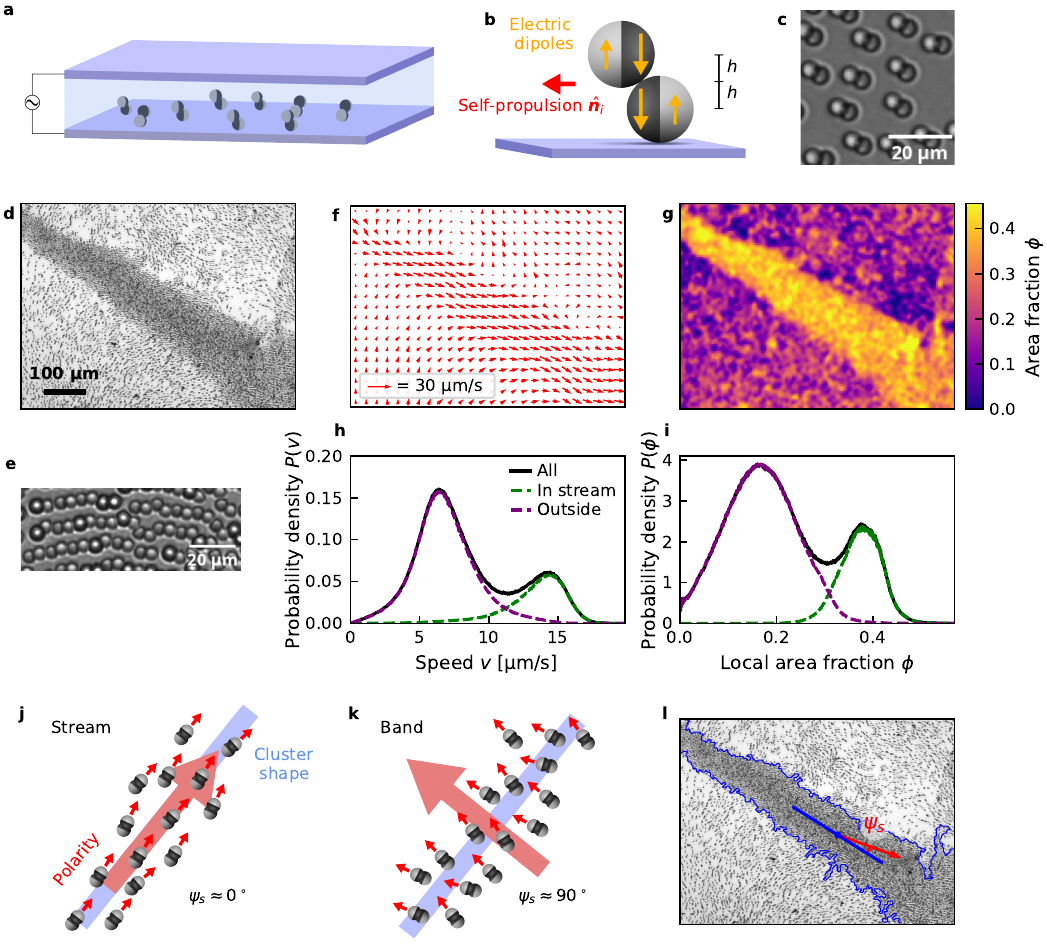}

    {\phantomsubcaption\label{fig:experiments-exptsketch}}
    {\phantomsubcaption\label{fig:experiments-dimersketch}}
    {\phantomsubcaption\label{fig:experiments-closeup}}
    {\phantomsubcaption\label{fig:experiments-stream}}
    {\phantomsubcaption\label{fig:experiments-chain}}
    {\phantomsubcaption\label{fig:experiments-flowfield}}
    {\phantomsubcaption\label{fig:experiments-areafrac}}
    {\phantomsubcaption\label{fig:experiments-velocity_distribution}}
    {\phantomsubcaption\label{fig:experiments-af_distribution}}
    {\phantomsubcaption\label{fig:experiments-streamsketch}}
    {\phantomsubcaption\label{fig:experiments-bandsketch}}
    {\phantomsubcaption\label{fig:experiments-streamangle}}
	\caption{
		\textbf{Active metal-dielectric Janus colloids dimerize and form polar streams.}
		(\subref*{fig:experiments-exptsketch}) Schematic of the experimental setup. Janus colloids sediment in deionized water in the sample cell, across which a vertical a.c.~electric field is applied. 
		(\subref*{fig:experiments-dimersketch}) Side-view sketch of a dimer. The external electric field induces electric dipoles of opposite orientation (orange arrows) and different magnitudes in the two hemispheres of each monomer. Electrostatic interactions between the dipoles in the metallic (black) hemispheres causes the colloids to dimerize. The dimers self-propel (red arrow) with the top monomer in front.  
		(\subref*{fig:experiments-closeup}) High-magnification top-view image of flocking dimers. The monomer on top (below) appears lighter (darker).
        (\subref*{fig:experiments-stream}) Microscope image of a stream of dimers in experiments (\hyperref[mov:expt_stream]{Movie 1}).
        (\subref*{fig:experiments-chain}) High-magnification top-view image of dimer chains.
        (\subref*{fig:experiments-flowfield},\subref*{fig:experiments-areafrac}) Velocity field (\subref*{fig:experiments-flowfield}) and area-fraction field (\subref*{fig:experiments-areafrac}) of the stream in (\subref*{fig:experiments-stream}).
        (\subref*{fig:experiments-velocity_distribution},\subref*{fig:experiments-af_distribution}) Distributions of dimer speed (\subref*{fig:experiments-velocity_distribution}) and local area fraction (\subref*{fig:experiments-af_distribution}) inside and outside the stream.
        (\subref*{fig:experiments-streamsketch},\subref*{fig:experiments-bandsketch}) Top-view sketches illustrating the difference between streams and bands: In streams (\subref*{fig:experiments-streamsketch}) particles move collectively (red) along the long axis of the dense cluster (blue). In bands (\subref*{fig:experiments-bandsketch}) particles move perpendicularly to the band's long axis. This difference can be quantified through the stream angle $\psi_\text{s}$, which is close to $0^\circ$ for streams and $90^\circ $ for bands. 
		(\subref*{fig:experiments-streamangle}) We measure the stream angle $\psi_\text{s}$ in experiments as the angle between the shape axis of a stream (blue) and its net velocity orientation (red). 
	}  
	\label{fig:experiments}
\end{figure*}

We prepare metal-dielectric Janus colloids~\cite{yan2016, zhang2021, das2024, vanderlinden2019, iwasawa2021, alvarez2026} by coating 3-$\mu$m-diameter silica spheres with 6 nm titanium and 5 nm silicon oxide on one hemisphere (\hyperref[sec:methods]{Methods}). The colloids are suspended in deionized water and placed between two conductive coverslips separated by a 120 $\mu$m spacer (\cref{fig:experiments-exptsketch}), where they sediment to form a monolayer on the bottom coverslip, covering area fractions of around $\phi_0 \approx 0.17-0.22$. We then drive the particles by applying a perpendicular a.c.~voltage with amplitude 10 V and frequency 30 kHz.

The resulting electric field causes the particles to turn their equator parallel to the field
and polarizes the two hemispheres of the colloids unequally due to their different dielectric properties \cite{gangwal2008}. Electrostatic interactions between the induced dipole moments cause the colloids to spontaneously dimerize (\cref{fig:experiments-dimersketch,fig:experiments-closeup}). Dimers form by one monomer going out of plane and on top of the other monomer such that the two metal-hemisphere dipoles are aligned \cite{chao2025}. A very small number of colloids remains monomeric. The dimers self-propel as indicated in \cref{fig:experiments-dimersketch}, and they interact with one another via electrostatic interactions. The self-propulsion is caused by electrohydrodynamic flows in the electric double layer surrounding the coverslip \cite{ristenpart2007,wang2014}.

Our experiments show that the colloidal dimers organize into polar streams: elongated high-density clusters with collective motion of dimers along the long axis of the stream (\cref{fig:experiments-stream}, \cref{extfig:expt_streams} and \hyperref[mov:expt_stream]{Movie 1}). A dense stream is surrounded by a dilute flocking phase, with which it exchanges particles. In our experiments, streams persist over a timescale of minutes and can be up to $\sim$1 mm in length and 100 $\mu$m wide. They typically flow into flocking regions with homogeneous high density, and, frequently, vortices (\hyperref[mov:expt_vortex]{Movie 2}).

In streams, Janus dimers frequently form single-file chains which can be tens of dimers long (\cref{fig:experiments-chain}). Chains inside streams frequently reorganize, breaking and reforming at different points. Small chains containing 3-5 dimers are also observed in the dilute flocking phase, but they are short-lived, disappearing after a few seconds.

To quantify the behavior of streams, we perform particle image velocimetry to measure the flow field $\vec v(\vec r,t)$ and evaluate the local area fraction $\phi(\vec r,t)$ for $4$ streams (\cref{fig:experiments-flowfield,fig:experiments-areafrac} and \hyperref[sec:methods]{Methods}). The colloids move faster by a factor of $1.95\pm 0.07$ inside streams compared to the dilute phase (\cref{fig:experiments-velocity_distribution}), despite the area fraction in streams being on average $2.33\pm 0.06$ times higher than outside (\cref{fig:experiments-af_distribution}), in contrast with the typically observed density-induced slowdown of non-aligning active particles \cite{tailleur2008,fily2012,bialke2013,cates2015}. The high speeds at high densities are enabled by the strong alignment of particles in the streams, which favors a fast and smooth flow. When moving in single-file chains, collisions that would cause a velocity decrease are less frequent.

The streams we observe differ from traveling bands commonly observed in systems of self-propelled particles close to the flocking transition \cite{chate2008,bricard2013,geyer2019,chate2020}. Both consist of an elongated high-density cluster of particles that move collectively. However, in streams, collective motion is along the long axis of the cluster (\cref{fig:experiments-streamsketch}), whereas a band travels perpendicularly to the extent of the cluster (\cref{fig:experiments-bandsketch}). To prove that what we observe are indeed streams and not bands, we segment the clusters in our experiments and compute the angle $\psi_\text{s}$ between their long axis and the average flow in the cluster (\cref{fig:experiments-streamangle} and \hyperref[sec:methods]{Methods}). This stream angle should be small for streams and around $90^\circ$ for traveling bands. We find $\psi_\text{s}=6.7^{\circ}\pm1.1^{\circ}$, indicating that our experiments exhibit streams --- a phenomenon distinct from traveling bands.

\section{Microscopic model of self-propelled Janus dimers}

\begin{figure}[t!]
	\centering
	\includegraphics[width=0.45\textwidth]{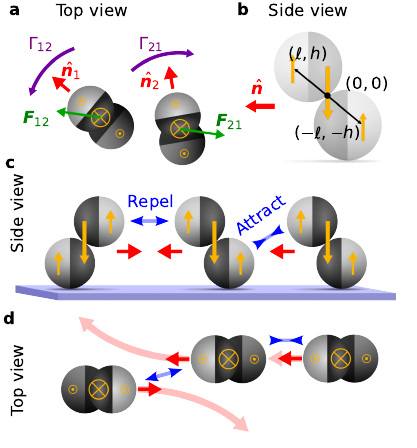}
    {\phantomsubcaption\label{fig:model-interactions_topview}}
    {\phantomsubcaption\label{fig:model-dipoles}}    
    {\phantomsubcaption\label{fig:model-interactions_sideview}}
    {\phantomsubcaption\label{fig:model-chain_formation}}
	\caption{
		\textbf{Model of Janus dimers.}
        (\subref*{fig:model-interactions_topview}) In our model, Janus dimers self-propel along their orientation ($\hat{\vec n}_i$). The electric dipoles (orange) induce pairwise forces $\vec F_{ij}$ and torques $\Gamma_{ij}$.
		(\subref*{fig:model-dipoles}) We model a Janus dimer using three electric dipoles (orange 
        arrows), two of which are horizontally ($\ell$) and vertically ($h$) displaced with respect to the center of mass (black point).
        (\subref*{fig:model-interactions_sideview}) Head and tail dipoles of two dimers do not sit in the same vertical plane and thus weakly attract each other, unlike in-plane interactions which are purely repulsive. 
        (\subref*{fig:model-chain_formation}) This arrangement makes the interactions between Janus dimers polar and favors chain configurations, since aligned particles in single file follow each other while anti-aligned particles scatter off one another.
	}  
	\label{fig:model}
\end{figure}

To investigate the formation of streams, we develop a microscopic model of self-propelled Janus dimers with electrostatic dipole-dipole interactions (\cref{fig:model-interactions_topview}). Each dimer $i$ is treated as a point particle at the dimer's center of mass $\vec r_i$ that propels with speed $v_0$ along $\hat{\vec n}_i(\theta_i)=(\cos\theta_i,\sin\theta_i)^T$, where $\theta_i$ is the particle orientation. Particles interact via pairwise forces $\vec F_{ji}$ and torques $\Gamma_{ji}$ and are subject to orientational noise with rotational diffusivity $D_\text{r}$. Translational noise is negligible in the experiments, so we do not include it in our model. In summary, our model is described by the Langevin equations
\begin{align}
    \frac{\dd\vec r_i}{\dd t}=&\,v_0\hat{\vec n}_i(\theta_i)+\frac{1}{\xi_\text{t}}\sum_{j\neq i}\vec F_{ji}, \label{eq:eom_f} \\
    \frac{\dd \theta_i}{\dd t}=&\,\frac{1}{\xi_\text{r}}\sum_{j\neq i} \Gamma_{ji}+\sqrt{2D_\text{r}}\eta_i(t), \label{eq:eom_t}
\end{align}
where we have also introduced the translational and rotational friction coefficients $\xi_\text{t}$ and $\xi_\text{r}$, respectively, and a Gaussian white noise $\eta_i(t)$ that satisfies $\langle \eta_i(t)\eta_j(t')\rangle=\delta_{ij}\delta(t-t')$.

As in previous work \cite{yan2016,zhang2021,das2024}, we assume interactions to be purely electrostatic. An order-of-magnitude estimate reveals hydrodynamic interactions to be negligible in comparison \cite{zhang2016}. Each monomer in a dimer has an electric dipole $\vec d_\text{d}$ in the dielectric hemisphere and a dipole $\vec d_\text{m}$ in the metal hemisphere. Since the two metal-hemisphere dipoles are one on top of the other in the dimer (\cref{fig:model-dipoles}), we replace them by a single effective dipole of strength $2\vec d_\text{m}$ for simplicity. Similarly, the head dielectric-hemisphere dipole is higher up than the tail one (\cref{fig:model-dipoles}), which breaks head-tail symmetry in the dimer interactions and makes them polar (\cref{fig:model-interactions_sideview,fig:model-chain_formation}). We provide estimates for the different parameters in \cref{si:parameters} and a detailed description of the interactions in \cref{si:model}.

\begin{figure*}[t!]
	\centering
	\includegraphics[width=1.0\textwidth]{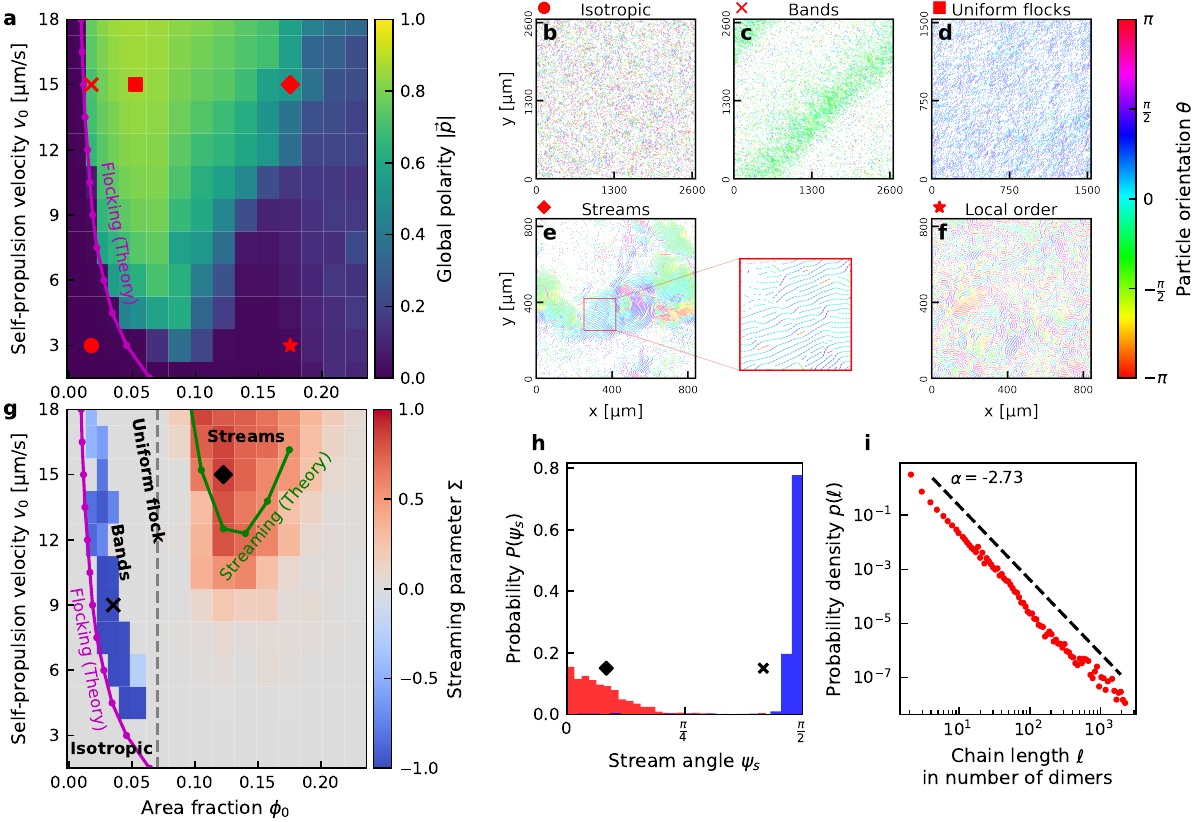}

    {\phantomsubcaption\label{fig:simulations-polarity}}
    {\phantomsubcaption\label{fig:simulations-isotropic}}
    {\phantomsubcaption\label{fig:simulations-bands}}
    {\phantomsubcaption\label{fig:simulations-flock}}
    {\phantomsubcaption\label{fig:simulations-stream}}
    {\phantomsubcaption\label{fig:simulations-localorder}}
    {\phantomsubcaption\label{fig:simulations-streamparam}}
    {\phantomsubcaption\label{fig:simulations-streamangledistribution}}
    {\phantomsubcaption\label{fig:simulations-chainlength}}
    
	\caption{
		\textbf{Flocking and stream phases of Janus particle dimers in simulations.} 
		(\subref*{fig:simulations-polarity}) Phase diagram showing global polarity, which reveals the flocking transition from an isotropic to a polar ordered state as area fraction or self-propulsion increase. The magenta curve is the flocking transition predicted from the hydrodynamic theory using the pair distribution function measured in simulations. 
		At different area fraction $\phi_0$ and self-propulsion speed $v_0$, we find 
		(\subref*{fig:simulations-isotropic}) a disordered isotropic state, 
		(\subref*{fig:simulations-bands}) traveling bands, 
		(\subref*{fig:simulations-flock}) uniform flocks, 
		(\subref*{fig:simulations-stream}) streams, which are made of single-file dimer chains, and 
		(\subref*{fig:simulations-localorder}) homogeneous states with local polar order. 
		(\subref*{fig:simulations-streamparam}) Phase diagram showing the time-averaged streaming parameter $\Sigma=\langle\cos(2\psi_\text{s})\rangle_t$, which distinguishes the stream phase ($\Sigma\approx +1$, red) from the traveling bands ($\Sigma\approx -1$, blue) occurring around the flocking transition. The green curve is the streaming transition predicted from the hydrodynamic theory. The magenta curve from \cref{fig:simulations-polarity} is shown again for reference. The grey dashed line indicates the area fraction at which we switch from a high to low clustering distance to identify bands and streams, respectively, as described in the \hyperref[sec:methods]{Methods}. 
        (\subref*{fig:simulations-streamangledistribution}) Distribution of the stream angle $\psi_\text{s}$ for a stream (red) and a band (blue). 
		(\subref*{fig:simulations-chainlength}) Chain lengths (in number of constituent dimers) are power-law distributed in the stream phase.
	}  
	\label{fig:simulations}
\end{figure*}

To study the emergent collective dynamics of our model, we perform Brownian dynamics simulations with $N=10^4$ particles in a square box of side $L$ with periodic boundary conditions (see \hyperref[sec:methods]{Methods}). We build a phase diagram (\cref{fig:simulations-polarity}) by varying the self-propulsion speed $v_0$ and the global area fraction $\phi_0=NA/L^2$, where $A$ is the area occupied by one dimer.

At low area fraction and self-propulsion, we find an isotropic state (\cref{fig:simulations-isotropic} and \hyperref[mov:sim_iso]{Movie 3}). As we increase either area fraction or self-propulsion speed, global polar order emerges, bearing the hallmarks of a flocking transition: Close to the onset of polar order, we observe the formation of traveling bands (\cref{fig:simulations-bands} and \hyperref[mov:sim_band]{Movie 4}), a well-known feature of flocks \cite{chate2008,bricard2013,geyer2019,chate2020}. Further from the transition line, the system reaches a homogeneous flocking state (\cref{fig:simulations-flock} and \hyperref[mov:sim_flock]{Movie 5}). The emergence of flocking is due to polar alignment between particles, which is enabled by the different vertical offsets $\pm h$ of the dielectric-hemisphere dipoles (\cref{fig:experiments-dimersketch}) as discussed above. If we set the vertical offset $h=0$ in simulations, the interactions have nematic symmetry, and the system forms neither flocks nor streams (\cref{sifig:nematic_interactions}).

As the area fraction is increased further from the flocking phase, the system forms streams (\cref{fig:simulations-stream}). They are made of several long single-file particle chains moving in parallel (\cref{fig:simulations-stream}, inset). Although streams have high local polar order, global polar order in the stream phase is lower than in the flocking phase since streams often bend. At intermediate area fractions ($\phi_0\approx0.1-0.14$), the dilute phase surrounding the streams is flocking (\hyperref[mov:sim_stream_flock]{Movie 6}), while at higher area fractions ($\phi_0\approx0.17-0.21$), the density of the dilute phase is so low that it becomes isotropic, with most particles absorbed into the stream (\hyperref[mov:sim_stream_iso]{Movie 7}).

Stream-like behavior, often referred to as polar bands or lanes, has been previously observed in simulations of self-propelled rods, mostly with soft volume-exclusion interactions \cite{wensink2012, abkenar2013,weitz2015,shi2018,grossmann2020, tanida2020, bar2020}. Our model does not include such interactions; in our simulations, electrostatic repulsion is too strong to allow for particle overlap, suggesting that a different mechanism is at play. Moreover, in both our experiments and simulations, streams emerge from flocks, which is not the case for the self-propelled rod systems.

Finally, we note that at high area fractions and low self-propulsion speeds, the system is uniformly dense and has no global order. Particles form shorter chains which coalesce into domains with local polar order (\cref{fig:simulations-localorder,sifig:global_local_order}).

As for our experimental data, we calculate the stream angle $\psi_\text{s}$ to identify streams and bands in our simulations. We identify high-density clusters of particles using a clustering algorithm, determine their long axis and compute the stream angle $\psi_\text{s}$ between this axis and the polarity vector of the clusters (see \hyperref[sec:methods]{Methods}). In \cref{fig:simulations-streamparam}, we plot the time-averaged value of $\Sigma=\cos(2\psi_\text{s})$, which we call the streaming parameter and takes values between $-1$ (band) and $+1$ (stream). Together with the distribution of the stream angle $\psi_\text{s}$ (\cref{fig:simulations-streamangledistribution}), the streaming parameter reveals the presence of both traveling bands close to the flocking transition as well as streams at high self-propulsion and area fractions (\cref{fig:simulations-streamparam}).

\section{Self-organization of active particles into mesoscopic chains}\label{sec:chains}

We now investigate the emergence of chains, which we observe in the streams both in experiments and simulations (\cref{fig:experiments-chain,fig:simulations-stream}). Chains form because aligned dimers repel each other less than anti-aligned dimers (\cref{fig:model-chain_formation}). This is due to a weak attractive interaction between the heads and tails of different dimers, caused by the different vertical positions of their head and tail monomers.

The uniform flock state in our simulations (\cref{fig:simulations-flock}) already contains small chains of particles. These chains are rarely more than a few particles long and short-lived: They are easily destroyed by collisions with other particles from the side. This is consistent with the particles in a chain not being bound: The interaction energy of two aligned dimers does not show a local minimum which could produce a bound state (\cref{sifig:dimer_int_energy}). However, in a stream composed of parallel-moving chains, the chains shield each other's flanks from sideway collisions with other particles that could break them, enabling the chain to grow further. This observation suggests that chain formation in our system is a genuinely many-body, non-equilibrium effect. Indeed, we find that the distribution of chain lengths follows a power law in the stream phase (\cref{fig:simulations-chainlength}), indicating that chains do not result from equilibrium polymerization kinetics, which would lead to an exponential distribution \cite{phillips2013}. 

Since the self-organization of active particles into large-scale assemblies is an ongoing subject of research \cite{huang2024a}, a further investigation of mesoscopic structures such as single-file chains in our system would be an interesting extension of our work.

\section{Hydrodynamic theory of flocks and streams}

To understand the origin of flocking and polar streams in our system, we coarse-grain our model into a hydrodynamic theory using the BBGKY hierarchy technique (\cref{si:coarsegrain}) \cite{bialke2013,speck2015,grossmann2020, speck2020, zhang2021,das2024}. We thus obtain hydrodynamic equations for the density field $\rho(\vec r,t)$, the polarization density $p_\alpha(\vec r;t)$ and the nematic order density $Q_{\alpha\beta}(\vec r;t)$. We provide the details of the derivation and the complete equations in \cref{si:coarsegrain}. Here, we present simplified equations where we neglect terms that are not relevant for the formation of flocks and streams.

Due to particle number conservation, the density equation takes the form of a continuity equation,
\beq \label{eq:density}
    \partial_t\rho=-\partial_\alpha J_\alpha, \quad J_\alpha= v_{\alpha\beta}[\rho,Q]p_\beta-D_{\alpha\beta}[\rho,Q]\partial_\beta \rho+..., \\
\eeq
where the current $J_\alpha$ has contributions from self-propulsion ($v_{\alpha\beta}$, \cref{eq:coeff_v_tens}) and diffusion ($D_{\alpha\beta}$, \cref{eq:coeff_D_tens}), which are both renormalized by the interactions, and can therefore depend on density and be anisotropic. For example, a decrease in the effective self-propulsion speed $v_{\alpha\beta}$ with density due to repulsive interactions is known to induce MIPS \cite{cates2015}.

The polarization dynamics take the form of Toner-Tu equations \cite{toner1995,toner2012,marchetti2013}, 
\begin{align}
    \partial_t p_\alpha +&\lambda_1 p_\beta\partial_\beta p_\alpha + \lambda_2(\partial_\beta p_\beta)p_\alpha + \lambda_3\partial_\alpha(p^2) \nonumber \\ =&\,  (a[\rho]-b|p^2|)p_\alpha-\partial_\alpha P[\rho] \nonumber\\ 
    &-\partial_\beta(w[\rho]Q_{\alpha\beta})-\nu Q_{\alpha\beta}\partial_\beta\rho-\gamma Q_{\alpha\beta} p_\beta+... \label{eq:polaritydyn_short}
\end{align}
The expressions for the coefficients are given in \crefrange{eq:coeff_lambda1}{eq:coeff_gamma}.
Going beyond the standard Toner-Tu equations, we include the dynamics of the nematic order field $Q_{\alpha\beta}(\vec r;t)$, which introduces new coupling terms. In particular, the term $-\partial_\beta(w[\rho]Q_{\alpha\beta})$ captures the polarity induced by inhomogeneities in the nematic order. Through the self-propulsion current in \cref{eq:density}, this polarity contribution gives rise to the so-called active nematic flux, which is responsible for phenomena like giant number fluctuations \cite{ramaswamy2003} and defect self-propulsion \cite{narayan2007} in dry active nematics. For $w[\rho]>0$, the active flux arises from extensile active nematic stresses.

Finally, for the dynamics of $Q_{\alpha\beta}(\vec r;t)$ we obtain
\beq \label{eq:nematicdyn_short}
    \partial_t Q_{\alpha\beta}=-4D_\text{r} Q_{\alpha\beta} + \gamma\Big(p_\alpha p_\beta-\frac{1}{2}\delta_{\alpha\beta}p^2\Big),
\eeq
where the first and second terms respectively describe the decay of nematic order due to orientational noise and the generation of nematic order associated with the presence of polar order. Our coarse-graining calculation (\cref{si:coarsegrain}) determines all hydrodynamic coefficients in the equations above in terms of microscopic parameters well as integrals of the pair distribution function $g(\vec r,\theta)$, which encodes the probability of finding a particle at a position $\vec r$ and orientation $\theta$ relative to a reference particle. To evaluate these integrals, we sample the pair distribution function numerically using our simulations (see \hyperref[sec:methods]{Methods} and \cref{sifig:pairdistr}). Thus, together with the expressions of their coefficients, \cref{eq:density,eq:polaritydyn_short,eq:nematicdyn_short} encode the hydrodynamic behavior of our system.

\begin{figure*}[t!]
	\centering
	\includegraphics[width=1.0\textwidth]{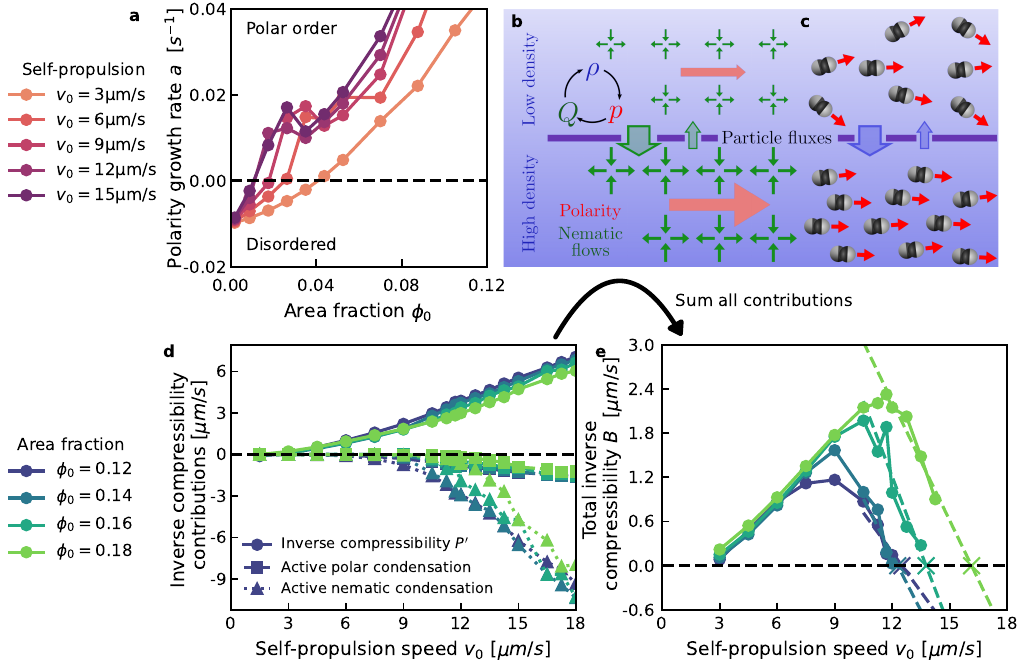}

    {\phantomsubcaption\label{fig:theory-apol}}
    {\phantomsubcaption\label{fig:theory-macro}}
    {\phantomsubcaption\label{fig:theory-micro}}
    {\phantomsubcaption\label{fig:theory-invcomp-contributions}}
    {\phantomsubcaption\label{fig:theory-invcomp}}
	\caption{
		\textbf{Stream formation is driven by active nematic condensation.} 
        (\subref*{fig:theory-apol}) Polarity growth rate $a$ as a function of area fraction $\phi_0$ for different self-propulsion speeds $v_0$. The flocking transition happens when $a$ turns positive. In the flocking phase, $a$ continues to increase with area fraction, since higher densities entail stronger alignment interactions.
        (\subref*{fig:theory-macro}-\subref*{fig:theory-micro}) Mechanisms of stream formation in a flock, from a macroscopic (\subref*{fig:theory-macro}) and a microscopic (\subref*{fig:theory-micro}) perspective.
        A feedback loop between density, polarity and nematic order field produces an instability that leads to stream formation. Macroscopically, denser regions entail higher polarity and hence higher nematic order, which drives a particle flux that further increases the local density. Microscopically, 
        the higher polarity in denser regions implies lower lateral outflux of particles towards regions of lower density, thus further increasing the local density.
        (\subref*{fig:theory-invcomp-contributions}) The different contributions to the effective inverse compressibility $B$ (\cref{eq:B}) as a function of self-propulsion speed $v_0$ for different area fractions $\phi_0$. The base inverse compressibility $P'$ of the system is always positive. However, the contributions of active polar condensation and active nematic condensation (terms proportional to $\lambda_3$ and $\gamma$ in \cref{eq:B}, respectively) reduce $B$, eventually causing it to become negative.
        (\subref*{fig:theory-invcomp}) Effective inverse compressibility $B$ as a function of self-propulsion speed $v_0$ for different area fractions $\phi_0$. When $B$ turns negative, the uniform flocking state becomes unstable. We determine the phase boundary (green line in \cref{fig:simulations-streamparam}) by extrapolating the values of $B$ in the flocking phase to $B=0$ (crosses), which marks the onset of the streaming instability. 
	}  
	\label{fig:theory}
\end{figure*}

\section{Flocking transition}

We now perform linear stability analysis of the hydrodynamic equations to predict the transitions from isotropic gas to flock (\cref{si:lsa}). The isotropic state with homogeneous density $\rho_0$ becomes unstable to polarity fluctuations if the polarity growth rate $a[\rho_0]$ becomes positive, which happens if torque-driven alignment of particles outcompetes rotational diffusion \cite{das2024}. We determine the onset of flocking by evaluating where $a[\rho_0]$ changes sign (\cref{fig:theory-apol}); this yields the magenta phase boundary in \cref{fig:simulations-polarity}, which is in good agreement with the simulations.

\section{Stream transition driven by active condensation}

Let us now consider the linear stability of the uniform flocking state. Here, the density is still homogeneous, $\rho(\vec r;t)=\rho_0$, but the system spontaneously breaks rotational symmetry to choose an axis $\hat{\vec x}$ to flock along, with polarization settling to its steady-state value 
\beq
p_\alpha(\vec r;t)=\sqrt{a[\rho_0]/(b+\gamma^2/8D_\text{r})}e_\alpha\equiv p_0e_\alpha, \label{eq:flock_polarity}
\eeq
where $e_x=1$ and $e_y=0$. A system with polar order also has nematic order; in our case, at steady state, it is given by 
\beq
Q_{\alpha\beta}(\vec r;t)=\frac{\gamma p_0^2}{4D_\text{r}}(e_\alpha e_\beta-\delta_{\alpha\beta}/2). \label{eq:flock_nematic}
\eeq
When considering perturbations around this steady state, we notice that there are only two slow modes: density perturbations $\delta\rho$, since they are associated with a conserved quantity, and orientation perturbations $\delta p_y$ of the global polarity vector, since they are a Goldstone mode. The other fluctuations, $\delta p_x$ and $\delta Q_{\alpha\beta}$, are fast modes that decay and become enslaved to the slow modes on timescales larger than $1/D_\text{r}$. After eliminating these fast modes, we assess the stability of the flock to perturbations ($\delta\rho,\delta p_y$) transverse to the flocking direction (given by \cref{eq:deltarho_linearized,eq:deltapy_linearized}), i.e. with a wavevector $\vec k\perp\hat{\vec x}$, which could produce lateral condensation of the flock into a stream. The growth rates of such perturbations have the form $\lambda_\pm=\pm ic_0|\vec k|$ (\cref{eq:evals_stream_instab}), which describe the propagation of sound modes with velocity $c_0=\sqrt{AB}$ in the flock \cite{tu1998,ramaswamy2010,toner2012,geyer2018}. Here, $A$ is a positive constant set by the hydrodynamic coefficients (\cref{eq:A}), and $B$ is an effective inverse compressibility for flocks (\cref{eq:B_si}) and reads
\beq
B=P'-\frac{\gamma p_0^2 (w'+\nu)}{8D_\text{r}}+\frac{a'p_0^2(2\lambda_3 -\gamma w/4D_\text{r})}{2a-\gamma^2p_0^2/4D_\text{r}},\label{eq:B}
\eeq
where primes denote derivatives with respect to the density, i.e. $a'=\dd a/\dd \rho|_{\rho_0}$ etc. Interestingly, $B$ can become negative, causing $c_0$ to become imaginary. This makes one of the modes unstable, leading to the formation of streams.

A useful comparison can be made with passive fluids, where spinodal decomposition occurs if the inverse compressibility $P'=\dd P/\dd \rho=\rho f''(\rho)$ is negative \cite{chaikin1995}. In these regions, the free energy density $f(\rho)$ is locally concave and thus a uniform state is unstable. MIPS of an isotropic active gas is also caused by a negative inverse compressibility derived from the active pressure \cite{takatori2014, solon2015a, obyrne2023} (see \cref{eq:effdiff,eq:invcomp}). Recent work has shown that chemotactic interactions between flocking particles can result in a negative inverse compressibility $P'$ and cause the formation of streams \cite{miller2024,miller2024a}. However, our analysis shows that streams can form even if $P'$ remains positive. Instead, an instability arises through the other terms in \cref{eq:B}, due to either of two effects, which we call active nematic and polar condensation, respectively.

Active nematic condensation is driven by the active nematic stresses $w[\rho]Q_{\alpha\beta}$, which are extensile in our system since $w[\rho_0]>0$ (\cref{eq:coeff_w}), and whose divergence appears in \cref{eq:polaritydyn_short}. These stresses (green arrows in \cref{fig:theory-macro}) are stronger in denser regions, since higher density leads to stronger polarity (\cref{eq:flock_polarity}), and hence stronger nematic order (\cref{eq:flock_nematic}). The resulting gradient in active stress causes a particle flux up the density gradient, perpendicular to the flocking direction, as illustrated in \cref{fig:theory-macro}. This mechanism therefore promotes the condensation of a flock into dense streams. A similar mechanism, albeit without involving polar order, can drive phase separation in active nematics \cite{bhattacharyya2023}.

Respectively, active polar condensation is based on the interplay between polarity and density through the term $\lambda_3\partial_\alpha(p^2)$ in \cref{eq:polaritydyn_short}, which can be interpreted as a polarity-dependent pressure \cite{toner1995}. In the parameter regime that we study, we find $\lambda_3<0$ (\cref{eq:coeff_lambda3}). Thus, the term $\lambda_3\partial_\alpha(p^2)$ generates a particle flux from regions of lower to higher polarity, which means towards denser regions since they develop a stronger polarity (\cref{eq:flock_polarity}). As sketched in \cref{fig:theory-micro}, this denser, more aligned region will have a small outflux of particles towards the less aligned region next to it. Conversely, the dilute, less aligned region will have a higher particle outflux towards the dense region, which again promotes an instability leading to the formation of streams.

As they are captured by the hydrodynamic equations, these two active condensation mechanisms are generic and should be present in any dry flock. Streams will form if these condensation effects are strong enough to outcompete pressure buildup in dense regions. Interestingly, while the $\lambda_3\partial_\alpha (p^2)$ term appears in the standard Toner-Tu equations \cite{toner1995,marchetti2013}, the active nematic term $-\partial_\beta(w[\rho]Q_{\alpha\beta})$ does not. This term scales like $\rho p_\alpha p_\beta$, and is thus of third order in the density and polarity fields, while the standard Toner-Tu equations include terms only up to second order in the fields.

To test which of these mechanisms is responsible for the streams observed in our system, we evaluate the inverse compressibility $P'$ as well as the effective inverse compressibility $B$; we use the expressions derived in our coarse-graining and the pair distribution functions measured in the simulations. We find that, in our system, $P'$ remains positive (\cref{fig:theory-invcomp-contributions}). However, active nematic condensation is sufficiently strong to overcome pressure and turn $B$ negative, causing the formation of streams (\cref{fig:theory-invcomp}). Active polar condensation occurs too, but it is too weak to lead to streams on its own (\cref{fig:theory-invcomp-contributions}). We thus conclude that the streams we observe in the Janus colloid dimer experiments form due to active nematic condensation. We predict the onset of streaming by determining where $B$ changes sign (\cref{fig:theory-invcomp}); this yields the green phase boundary in \cref{fig:simulations-streamparam}, which captures the behavior seen in simulations.

\section{Discussion and outlook}

To summarize, in experiments with dimerized Janus colloids, we showed that flocks can condense into polar streams --- elongated dense clusters with collective motion along their long axis. We developed and simulated a microscopic active particle model for the Janus dimers, which gives rise to a variety of states including uniform flocks, well-known transverse bands, and the polar streams. The simulations also reproduce the mesoscopic single-file particle chains that we observed in the experiments. We then systematically coarse-grained the model to derive hydrodynamic equations; this revealed two generic mechanisms --- which we called active nematic and polar condensation --- for the formation of polar streams.

Polar streams were recently predicted to emerge in flocks where particles interact through a chemical field that they produce \cite{miller2024,miller2024a}. Here, we demonstrate polar streams experimentally and we identify two additional mechanisms whereby they can emerge. These mechanisms do not rely on a mediating chemical field, and they should operate in any flock where particles repel. Therefore, our findings reveal general ways in which flocks can be shaped. In our system, we found that streams emerge mostly through the active nematic condensation mechanism. This finding highlights the relevance of nematic order in shaping flocks, despite it often being neglected in theories of flocking, including the standard Toner-Tu equations \cite{toner1995,marchetti2013}.

Phenomena resembling polar streams have been observed in simulations of various active particle models, including self-propelled rods with soft repulsion \cite{abkenar2013,weitz2015,shi2018,grossmann2020,wensink2012,bar2020}, particles with both turn-towards and alignment torques \cite{shea2025} or with contact-following interactions \cite{hiraiwa2020}, and active Potts models \cite{mangeat2020}. Polar clusters, albeit not as elongated as the streams, have also been found in systems of self-propelled particles with repulsion and polar alignment \cite{martin-gomez2018}. In these instances, the underlying hydrodynamic mechanisms remain unclear. Our work opens the door to investigating whether such polar clusters arise from the active condensation effects reported here. Other mechanisms are also conceivable: Notably, polar clusters might form via MIPS of the isotropic state. Subsequently, in the dense phase, alignment interactions could be sufficiently strong to create polar order. This scenario, with phase separation followed by flocking, would be the inverse of what we found, where the system first flocks and then phase separates.

Phase separation within a flock was previously found in Quincke rollers, where the onset of MIPS led to a loss of polar order \cite{geyer2019}. Our work reveals a different kind of phase separation of flocks in which collective motion is not only sustained but even enhanced: The formation of streams, and the mesoscopic particle chains within them, reduce interparticle collisions that would slow down particle motion. As a result, particles in the dense streams move faster than in the more dilute flocks surrounding them (\cref{fig:experiments-velocity_distribution}).

Therefore, we speculate that polar streams might be a collective state that allows for structured and efficient motion and transport. This idea could be relevant for stream-like behaviors found in biological systems, such as embryonic neural crest cells \cite{szabo2019,weissenbruch2025}, ant trails \cite{couzin2003}, and wildebeest herds \cite{wu2023}. Furthermore, contact following interactions in collective cell migration \cite{li2018,fujimori2019,alert2020} are similar to those leading to chain formation in our system. Interestingly, simulations of self-propelled particles that include such interactions show stream-like states \cite{hiraiwa2020}. Investigating stream and chain formation in biological active matter remains an exciting challenge for future work.

\bigskip

\textbf{Acknowledgments.}
We thank Saraswat Bhattacharyya, Matteo Ciarchi, Suchismita Das, Sagnik Garai, Frank J\"{u}licher, Roderich Moessner, Ignacio Pagonabarraga and John Toner for inspiring discussions. We thank Steve Granick for agreeing to share the experimental data. We acknowledge computing support by the computing facility at MPI-PKS.

\bigskip
\textbf{Funding statement.}
X.D., Z.Z. and J.Z. acknowledge financial support from the National Natural Science Foundation of China (NSFC) (Grant No. T2522036, 22273099), Chinese Academy of Sciences (KJ2060007008) and University of Science and Technology of China (KY2060000273 and KY2060000211). X.D., Z.Z. and J.Z. acknowledge the Supercomputing Center of University of Science and Technology of China (USTC) for providing computational resources. The experimental work was partially carried out at the Center for Micro and Nanoscale Research and Fabrication and the Instruments Center for Physical Science, University of Science and Technology of China.
R.A. acknowledges funding from the European Union through ERC Starting Grant ``Living\_Fluctuations'' (No. 101114584).
M.B. was supported by the European Union (ERC, QuSimCtrl, 101113633). Views and opinions expressed are, however, those of the authors only and do not necessarily reflect those of the European Union or the European Research Council Executive Agency. Neither the European Union nor the granting authority can be held responsible for them. 

\bigskip

\textbf{Author contributions.} Z.Z. and J.Z. performed experiments. L.H., X.D. and J.Z. analyzed experimental results. L.H., M.B. and R.A. developed the model. L.H. performed and analyzed the simulations and the theory calculations. M.B. and R.A. supervised the work. L.H., M.B., and R.A. wrote the paper, which was reviewed by all authors.

The authors used Claude Opus 5.5 and GPT-5.3-Codex during the preparation of this manuscript to assist with code optimization as well as for plotting and documentation purposes. 
The authors assume full responsibility for the scientific content, accuracy, and integrity of this paper.

\bigskip

\textbf{Data availability statement.}
The data and code associated with this manuscript are available under DOI:10.5281/zenodo.23953164.

\bibliography{references}

\onecolumngrid

\clearpage

\twocolumngrid

\begin{appendix}

\setcounter{figure}{0}                          
\renewcommand{\thefigure}{\arabic{figure}}      
\renewcommand{\theHfigure}{ED.\arabic{figure}}  
\renewcommand{\figurename}{Extended Data Fig.}  

\crefalias{figure}{extfigure}   
\crefalias{subfigure}{extsubfigure}

\section{Methods} \label{sec:methods}

\subsection{Experiments}

\subsubsection{Particle Synthesis}
Following protocols described in Ref.~\cite{yan2016}, a submonolayer of 3-$\mu$m-diameter silica particles (Tokuyama) is prepared on a standard glass slide. Then, using an electron beam evaporator, 6 nm of titanium and then 5 nm of silicon oxide (SiO$_2$) are deposited vertically on the glass slide. The preparation is then washed with isopropyl alcohol and deionized water and finally sonicated into deionized water to collect the Janus particle monomers.

\subsubsection{Experimental Setup}
Two coverslips (SPI Supplies) are coated with indium tin oxide to make them conductive, followed by 25 nm of silicon oxide to prevent particles from sticking to them. The coverslips have a 9 mm hole in the center, separated by a 120-$\mu$m-thick spacer (GraceBio SecureSeal), into which we place the Janus particle suspension.

An alternating voltage is applied between the coverslips using a function generator (Agilent 33522A). Microscopic images and videos are taken with a CMOS camera (Edmund Optics 5012M GigE) with 20 ms time resolution. The sample cell is imaged with 10$\times$ and 20$\times$ air objectives on an inverted microscope (Axiovert 200). The observation areas are 768 $\mu$m $\times$ 576 $\mu$m, and 384 $\mu$m $\times$ 288 $\mu$m, respectively.

\subsubsection{Image Analysis}
Image processing is performed using MATLAB with home-developed codes.

\subsubsection{Experimental Data Analysis}
Due to their large size, we can only image streams with low magnification (10$\times$). At this magnification, we cannot resolve single particle positions and orientations, in particular not inside the very dense streams. Instead, we use particle image velocimetry to obtain the flow field $\vec v(\vec r,t)$, which was previously found to be a good proxy for particle orientation in similar experiments \cite{das2024}. We also estimate for the local area fraction $\phi(\vec r,t)$ by binarizing the image, dividing it into a grid of size 159 $\times$ 119, and dividing the number of occupied pixels in each grid cell by the pixel count of an isolated dimer.

We identify high-density clusters in each field of view by thresholding the area fraction field at 1.5$\times$ its time-averaged value. In the resulting binarized data, we determine the connected components and fill in small holes. This then serves as a segmentation mask for high-density clusters.

We then fit an ellipse to the mask of each cluster to get the orientation of its long axis. We also apply the mask to the flow field $\vec v(\vec r,t)$ to get the flow field inside the cluster, of which we obtain the average orientation as a proxy for cluster polarity. We then define the stream angle as the angle between the long axis of the fitted ellipse and the average velocity vector.

\subsection{Simulations}

\subsubsection{Simulation scheme}
We perform Brownian dynamics simulations of \cref{eq:eom_f,eq:eom_t} to simulate the collective behavior of $N=10^4$ Janus colloid dimers in a square box of side length $L$ with periodic boundary conditions. We use the simulation units and parameter values specified in \cref{table:params_sim} in the Supplemental Material. We use an explicit Euler-Maruyama method with time step $\Delta t=10^{-5}$ for time evolution, and run each simulation for a total time of $T=150$ simulation time units, corresponding to 4 hours and 10 minutes.

For better performance, we truncate interactions at a cutoff distance $r_\text{cut}=8$. We use a cell-list algorithm to track particle neighbors. In our simulations, we vary the total area fraction $\phi_0=NA/L^2$ by adjusting the simulation box size $L$, keeping $N$ and the particle area $A$ fixed. We vary the area fraction $\phi_0$ and self-propulsion speeds $v_0$ in the ranges $\phi_0\in[0.002,0.23]$ and $v_0\in[0,18]$ $\mu$m/s. We benchmark our code using results from \cite{zhang2021,das2024} by using the models from these articles. 

\subsubsection{Pair distribution function}
We use our simulations to numerically evaluate the pair distribution function $g(r,\varphi,\theta)$, which is a function of the relative particle positions in polar coordinates $(r,\varphi)$ and the relative particle orientation $\theta$. Once the simulations with $N=10000$ particles have reached steady state, we build a histogram $\mathcal N(r,\varphi,\theta)$, using bin sizes $\Delta r=0.1$ and $\Delta\varphi=\Delta\theta=\pi/180$, and average over $t_\text{run}=50$ snapshots. We then compute the pair distribution function as in Ref.~\cite{das2024}, using
\beq
g(r,\varphi,\theta)=\frac{2\pi \mathcal N(r,\varphi,\theta)}{A(r)N t_\text{run}\rho\Delta\theta},
\eeq
where $A(r)=r\Delta r\Delta\varphi$ is the area of the annular segment of radial width $\Delta r$ and angular opening $\Delta \varphi$ at radial distance $r$. 

\subsubsection{Determining particle cluster geometry}
In our simulations, we quantify the tendency of the active particles to form streams by identifying high-density clusters of particles and comparing the shape of a cluster with its net polarity. We identify two particles as belonging to the same cluster if they are closer than some cutoff distance $r_\text{cl}$.

We choose $r_\text{cl}$ based on the shape of the pair distribution function $g(r,\varphi,\theta)$, as shown in \cref{sifig:streamparam_cutoffs}. In the stream phase, $g$ shows clear peaks along the reference particle axis ($y=0$) and perpendicular to it ($x=0$), which inform us about the typical distance between particles in a chain and between chains forming a stream, respectively (see \cref{sifig:streamparam_cutoffs-gstream}). By choosing a cutoff distance slightly above these values, the clustering algorithm identifies groups of particles belonging to the same chain or stream.

However, in the traveling bands observed around the flocking transition, interparticle distances are much larger since this transition happens at lower area fractions. Using the clustering algorithm with the cutoff distance obtained from the pair distribution function in the stream phase will therefore not recognize any bands. Hence we use a larger $r_\text{cl}$ for low area fractions ($\phi\leq0.07$), chosen around the minimum after the main peak in the radial distribution function in the band-forming regime (see \cref{sifig:streamparam_cutoffs-gband}). The results of our analysis for either choice of $r_\text{cl}$ across all conditions is shown in \cref{sifig:streamparam_cutoffs-streamparam_low} and \cref{sifig:streamparam_cutoffs-streamparam_high}.

Once we have identified clusters, we analyze their shape. This requires careful treatment of the periodic boundary conditions.  First, we construct the adjacency graph of points in a cluster. Then, starting from a reference particle, we perform a breadth-first traversal of the adjacency graph, converting each particle position to the minimum image of the reference particle.

Once the entire cluster has been unwrapped, we fit an ellipse to the point cloud of particle positions and identify the long axis of the cluster. We then calculate the angle between the long axis of the fitted ellipse and the cluster polarity orientation, which gives us the stream angle $\psi_\text{s}$. We only count clusters that contain at least 5\% of all particles (in order to avoid picking up short-lived local fluctuations of the density) and whose average area fraction exceeds the global area fraction $\phi_0$ by at least 20\% (in order to avoid identifying large chunks of a homogeneous state as a potential stream or band).

To obtain the values plotted in \cref{fig:simulations-streamparam}, we average $\cos(2\psi_\text{s})$ of the largest cluster over time, and set this value to zero for a given time point if no eligible cluster is present.

\onecolumngrid

\cleardoublepage

\twocolumngrid

\begin{figure*}
	\centering
	\includegraphics[width=0.98\textwidth]{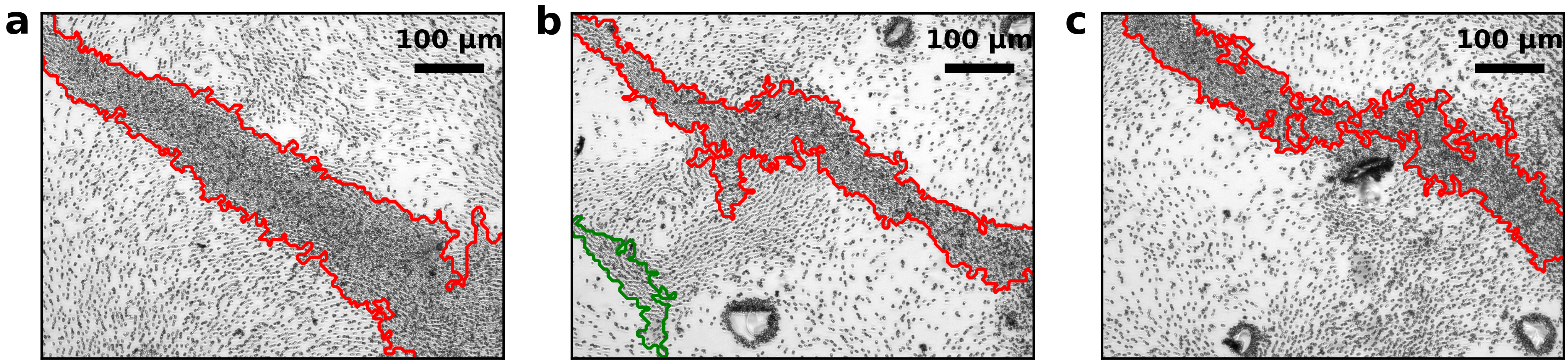}
	\caption{
		\textbf{Examples of streams in Janus colloid dimer experiments.}
        Here, we show other instances of streams observed in the experiments with dimerized Janus colloids, and highlight them using the segmentation masks obtained from the area-fraction field.
    	}  
	\label{extfig:expt_streams}
\end{figure*}

\begin{figure*}
	\centering
	\includegraphics[width=0.5\textwidth]{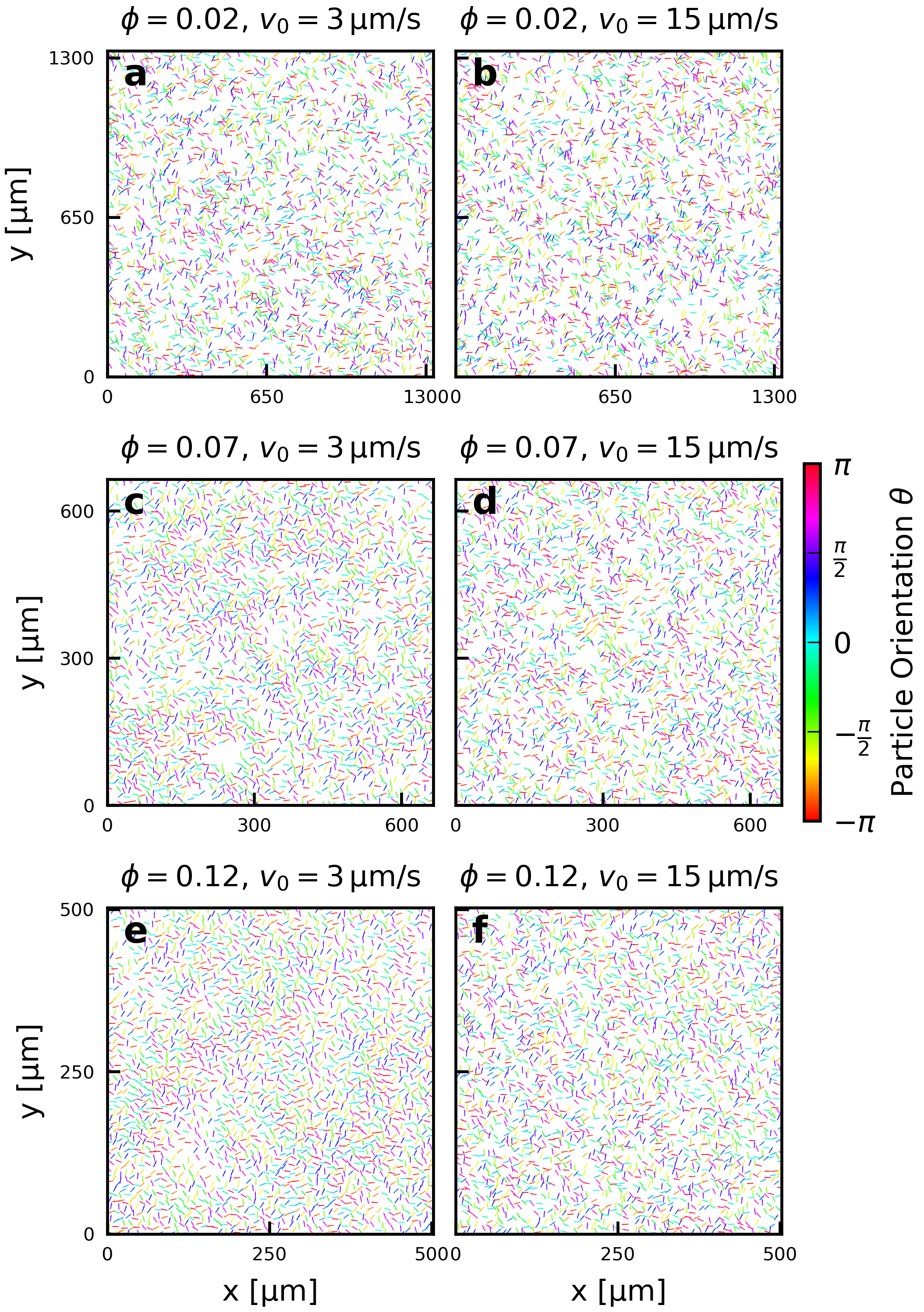}
	\caption{
		\textbf{Head-tail asymmetric interactions are necessary for flocking.} 
		To demonstrate the importance of the vertical offset $2h$ between the head and tail monomers (\cref{fig:experiments-dimersketch}), we simulate our microscopic model with $h=0$. This makes interactions head-tail symmetric. In this case, neither flocks nor streams are observed and the system remains isotropic.
	}  
	\label{sifig:nematic_interactions}
\end{figure*}

\begin{figure*}
	\centering
	\includegraphics[width=0.6\textwidth]{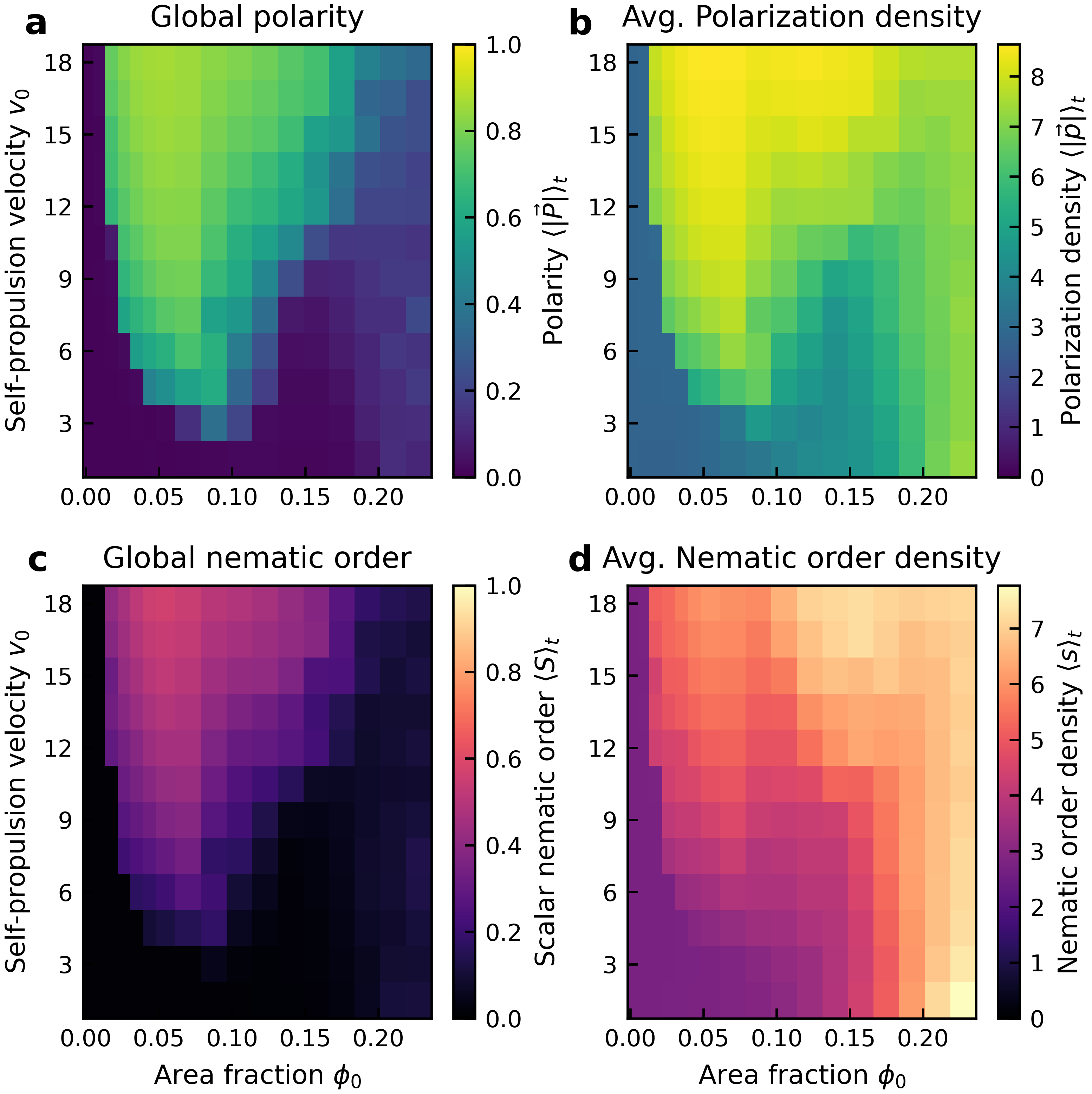}
    {\phantomsubcaption\label{sifig:global_local_order-pglobal}}
    {\phantomsubcaption\label{sifig:global_local_order-plocal}}
    {\phantomsubcaption\label{sifig:global_local_order-sglobal}}
    {\phantomsubcaption\label{sifig:global_local_order-slocal}}
	\caption{
		\textbf{Global and local order.}
		We compare global and local in our system by plotting the time-averaged (\subref*{sifig:global_local_order-pglobal}) global polarity and (\subref*{sifig:global_local_order-plocal}) local polarization density, and similarly (\subref*{sifig:global_local_order-sglobal}) global nematic order and (\subref*{sifig:global_local_order-slocal}) the local nematic order density. Even though global order decreases at high area fractions, in particular at low self-propulsion speeds, average local order remains high or even increases. This is because streams have a high local polarization but may bend, which decreases global polarity.
	}  
	\label{sifig:global_local_order}
\end{figure*}

\begin{figure*}
	\centering
	\includegraphics[width=0.6\textwidth]{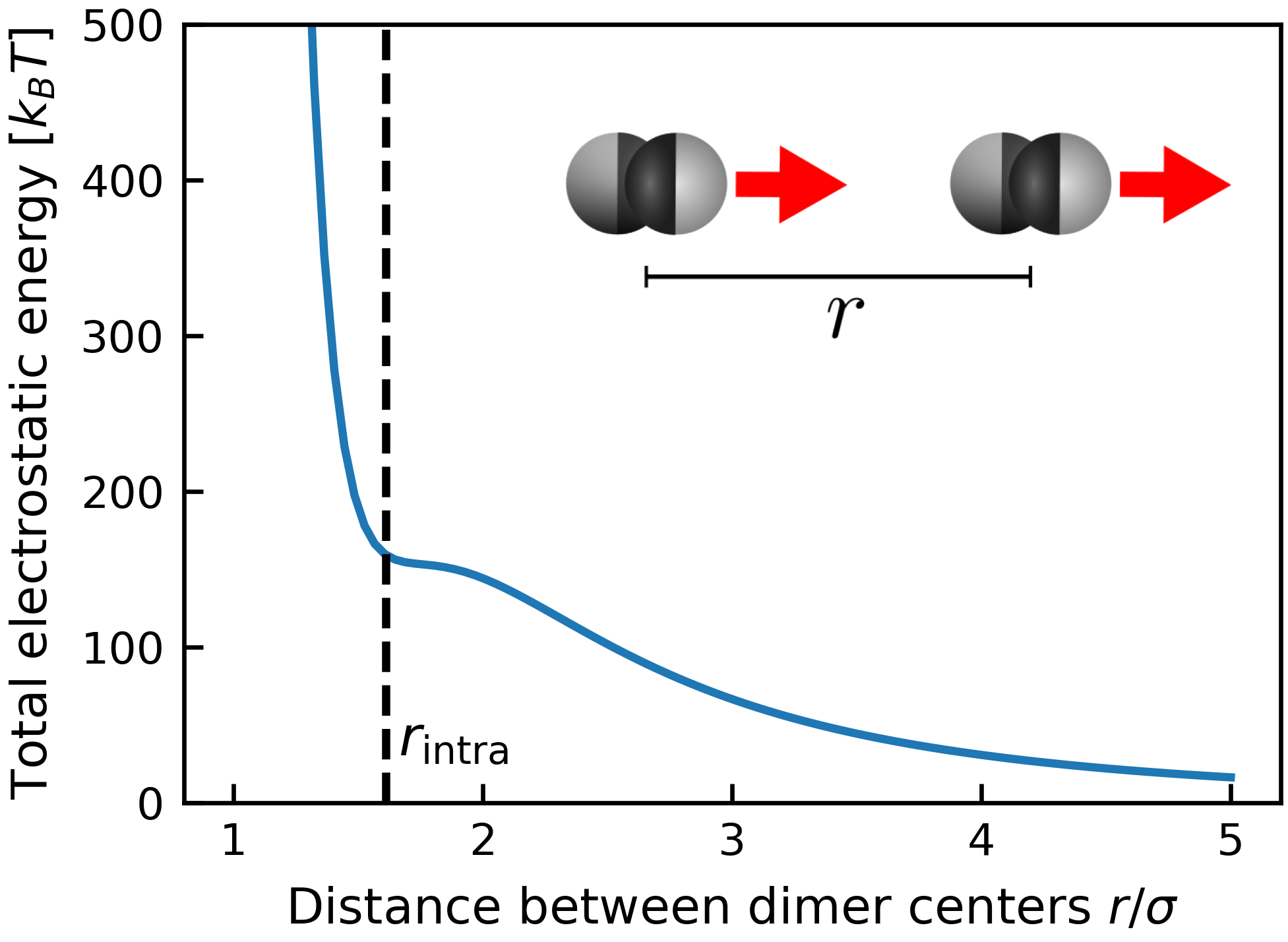}
	\caption{
		\textbf{Chains are not a bound state of dimers.}
		We plot the interaction energy (due to electric dipole-dipole interactions) between two aligned dimers as a function of the distance between them, as shown in the inset sketch. Since no local minimum exists, there is no bound state protected by an energy barrier. However, there is a plateau in the interaction energy which coincides with the typical distance between particles in a chain (see \cref{fig:theory,sifig:streamparam_cutoffs}). 
	}  
	\label{sifig:dimer_int_energy}
\end{figure*}

\begin{figure*}
	\centering
	\includegraphics[width=0.98\textwidth]{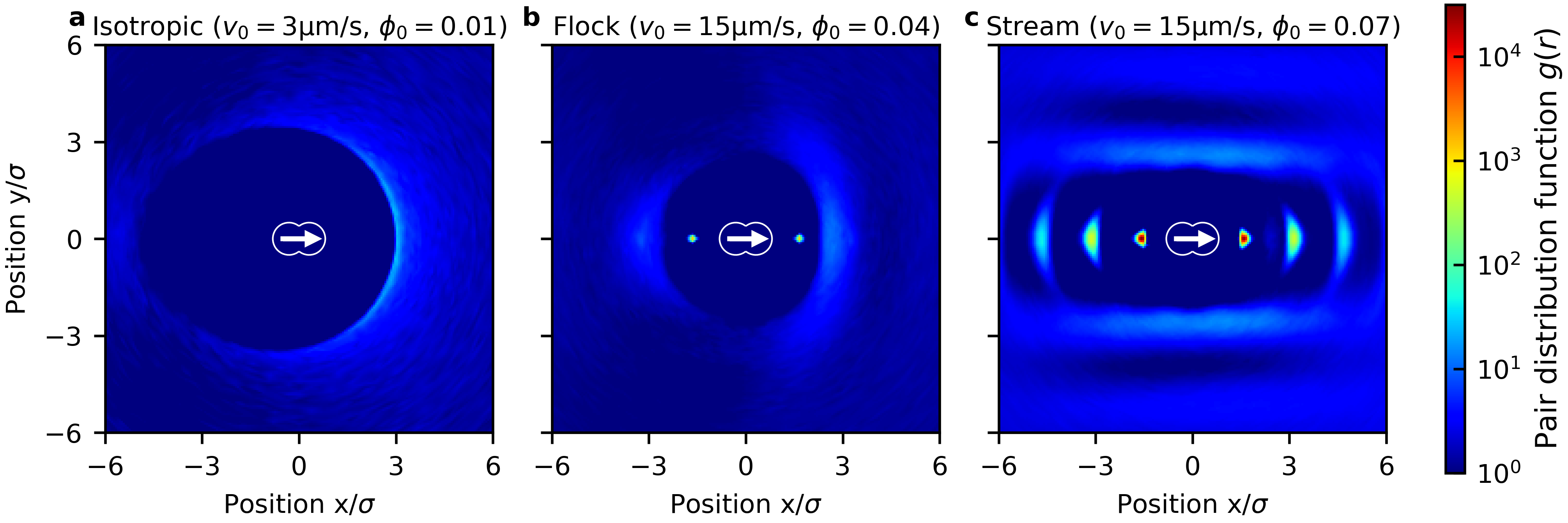}
    {\phantomsubcaption\label{sifig:pairdistr-giso}}
    {\phantomsubcaption\label{sifig:pairdistr-gflock}}
    {\phantomsubcaption\label{sifig:pairdistr-gstream}}
	\caption{
		\textbf{Pair distribution function.}
		The coefficients of our hydrodynamic theory are determined by the particle interactions and correlations. We obtain the correlations by sampling the pair distribution function $g(\vec r,\theta)$ in our simulations. Here, we show $g$ after integrating out the relative orientation $\theta$,
        (\subref*{sifig:pairdistr-giso}) in the isotropic phase, 
        (\subref*{sifig:pairdistr-gflock}) in the flocking phase, and 
        (\subref*{sifig:pairdistr-gstream}) in the stream phase.
		The white arrow indicates the orientation of the reference dimer, whose shape is shown for scale. Two peaks in front and behind the reference dimer are already visible in the flocking phase, pointing to incipient chain formation. In the stream phase, further peaks in front and behind the reference particle become visible, indicating the formation of mesoscopic multi-particle chains. Neighboring chains, which together form streams, are visible as smeared out stripes parallel to the reference dimer.
	}  
	\label{sifig:pairdistr}
\end{figure*}

\begin{figure*}
	\centering
	\includegraphics[width=0.48\textwidth]{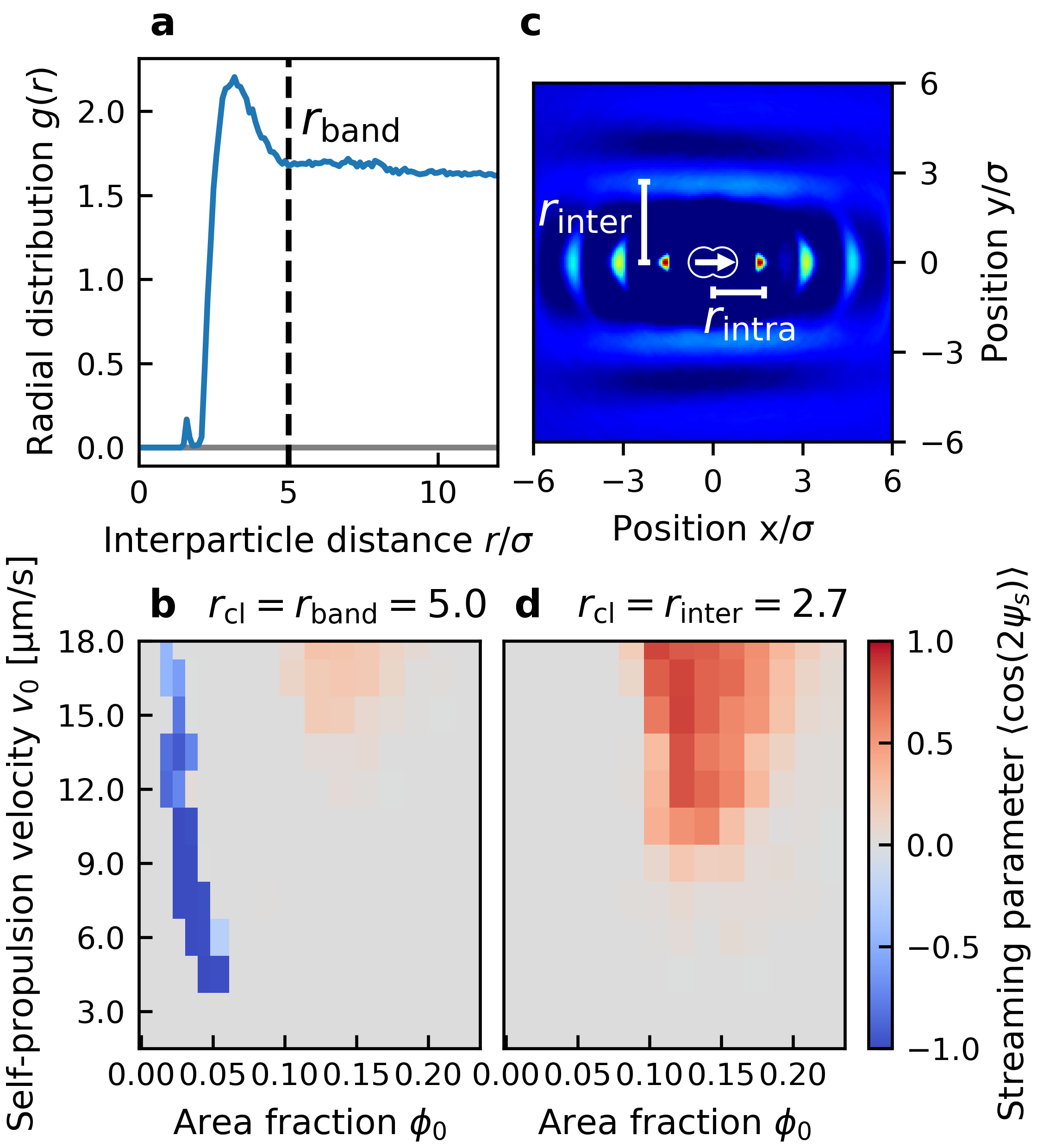}
    {\phantomsubcaption\label{sifig:streamparam_cutoffs-gband}}
    {\phantomsubcaption\label{sifig:streamparam_cutoffs-streamparam_low}}
    {\phantomsubcaption\label{sifig:streamparam_cutoffs-gstream}}
    {\phantomsubcaption\label{sifig:streamparam_cutoffs-streamparam_high}}
	\caption{
		\textbf{Stream parameter for different choices of cluster distance cutoff.}
		To compute the stream parameter $\langle\cos(2\psi_\text{s})\rangle$ shown in \cref{fig:simulations-streamparam}, we group particles in our simulations into clusters. Two particles separated by distance $r$ are considered to be in the same cluster if $r<r_\text{cl}$. We use two different values for the clustering cutoff $r_\text{cl}$, for low and high global area fractions $\phi_0$. 
        (\subref*{sifig:streamparam_cutoffs-gband}) We obtain $r_\text{cl}=5.0$ as cutoff for low area fractions by finding the minimum after the primary peak in the radial distribution function $g(r)$ in the band-forming phase (here $\phi_0=0.02$ and $v_0=15$ $\mu$m/s). 
        (\subref*{sifig:streamparam_cutoffs-streamparam_low}) When using this cutoff distance, we correctly identify the traveling bands appearing around the flocking transition, with $\langle\cos(2\psi_\text{s})\rangle\approx -1$. However, at high area fractions, using a large cutoff indiscriminately identifies large regions of the systems as high-density-clusters and thus cannot properly identify streams. 
        (\subref*{sifig:streamparam_cutoffs-gstream}) We therefore use the pair distribution function in the stream phase (here $\phi_0=0.12$ and $v_0=15$ $\mu$m/s) to identify the distance between neighboring chains in a stream and use this as clustering cutoff distance, $r_\text{cl}=r_\text{inter}=2.7$. We also indicate how we use the pair distribution function to measure the distance between dimers in the same chain ($r_\text{intra}=1.7$), which we use to identify which particles belong to the same chain (\cref{fig:simulations-chainlength}).
        (\subref*{sifig:streamparam_cutoffs-streamparam_high}) This allows us to correctly identify the formation of streams with $\langle\cos(2\psi_\text{s})\rangle\approx+1$. However, this cutoff distance is too small to capture the traveling bands that occur at lower densities.
	}  
	\label{sifig:streamparam_cutoffs}
\end{figure*}

\onecolumngrid

\end{appendix}


\resumetoc

\cleardoublepage

\renewcommand{\thefigure}{S\arabic{figure}}
\setcounter{figure}{0}
\counterwithout{equation}{section}
\renewcommand{\theequation}{S.\arabic{equation}}
\setcounter{equation}{0}
\renewcommand{\thesection}{S\arabic{section}}
\setcounter{section}{0}

\onecolumngrid
\begin{center}
	\textbf{\large{\textit{\nqdcolor{Supplemental Material}}}}\\
	\hfill \break
\end{center}
\twocolumngrid


\tableofcontents

\newpage

\section{Parameter estimates} \label{si:parameters}

Here, we determine the values of the physical parameters relevant for the experiments, which we then use in our simulations, including the magnitudes and positions of the electric dipoles in the Janus colloids as well as translational and rotational friction coefficients of the Janus colloid dimers. All parameters are summarized in \cref{table:params}, including their values in simulation units.

\begin{table*}[t]
\centering
\caption{\textbf{Parameter estimates and simulation units, based on Ref.~\cite{zhang2021}.}}
\label{table:params}

\begin{tabular}{lcc}
 \hline
 Description and symbol & Estimated physical value & Value in simulation units \\
 \hline\hline
 Colloid diameter $\sigma=2R$  & $3$ $\mu$m  & 1  \\
 Dimer rotational diffusion coefficient $D_\text{r}$ & 0.01 s$^{-1}$ & 1 \\
 Thermal energy $k_B T$ & $4.11\times 10^{-21}$ J & 1 \\
 Dielectric permittivity of the solvent $\epsilon_\text{w}$ & $6.95\times 10^{-10}$ C$^2$/(N m$^2$) & 1 \\
 Dimer translational friction coefficient & 32.5 mPa s $\mu$m & 0.71 \\
 Dimer rotational friction coefficient & 330 mPa s $\mu$m$^3$ & 0.80 \\
 Distance between dipoles $\ell$ & 1.5 $\mu$m & 0.5 \\
 Screening length $\lambda$ & 120 $\mu$m & 40 \\ 
 Vertical dipole offset $2h$ & 2.49 $\mu$m & 0.83 \\ 
 Dielectric dipole magnitude $|\vec d_\text{d}|$ & $4.64\times 10^{-22}$ C m & 52.84 \\
 Metal dipole magnitude $|\vec d_\text{m}|$  & $1.18\times 10^{-21}$ C m & 134.36 \\

\hline
\end{tabular}
\label{table:params_sim}

\end{table*}

\subsection{Dielectric response of Janus colloids}
The Janus particles are spheres made of silica with $2R=3$~$\mu$m diameter. One hemisphere has been coated with $d_\text{Ti}=6$~nm titanium, after which a $d_\text{SiO$_2$}=5$~nm protective layer of silicon oxide is added. The alternating voltage (amplitude $V=10$~V, frequency $\nu=30$~kHz) which is applied to the electrodes creates an electric field $\vec E_0$ which induces an electric polarization in each hemisphere. Given the $120$~$\mu$m separation between the electrodes, this field has a magnitude $E_0=83$~V/mm.

The induced complex dipole moment of the metal (dielectric) hemisphere $\vec d_\text{m/d}^c$ is proportional to the applied electric field,
\beq
\vec d_\text{m/d}^c(\nu)=\alpha_\text{m/d}(\nu)\vec E_0,
\eeq
where $\alpha_\text{m/d}(\nu)$ is the frequency-dependent complex polarizability of the corresponding hemisphere, which we assume to be half the polarizability of a full sphere made of the same material \cite{jones1995}:
\beq
\alpha_\text{m/d}(\nu)=2\pi\epsilon_\text{w} K_\text{m/d}(\nu)R^3,
\eeq
with $\epsilon_\text{w}$ being the dielectric permittivity of the surrounding water. The Clausius-Mossotti function $K_\text{m/d}(\nu)$ describes the frequency-dependent response of the metal (dielectric) hemisphere, which is caused by the dynamic response of the electrons in the material (Ohmic losses) and the Maxwell-Wagner effect at the interface of materials with different dielectric properties. These cause a phase lag in the induced dipole moments with respect to the driving field. However, we are only interested in time-averaged interactions between particles since the a.c. frequency $\nu$ is large compared to all other timescales in the system. We may thus neglect the phase shift and use an effective polarizability
\beq
\alpha_\text{m/d}(\nu)\approx2\pi\epsilon_\text{w} |K_\text{m/d}(\nu)|R^3.
\eeq

We proceed now to calculate the Clausius-Mossotti function for both hemispheres. These were also previously measured for the Janus particles used in our experiments \cite{yan2016}.

\subsubsection{Polarizability of the dielectric hemisphere}
For the dielectric hemisphere, we use the standard Clausius-Mossotti expression \cite{jones1995,shilov2001}
\beq
K_\text{d}(\nu)=\frac{\epsilon^c_\text{s}(\nu)-\epsilon^c_\text{w}(\nu)}{\epsilon^c_\text{s}(\nu)+2\epsilon^c_\text{w}(\nu)},
\eeq
where $\epsilon_{\text{s/w}}^c$ is the complex permittivity of silica (water):
\beq
\epsilon_{\text{s/w}}^c(\nu)=\epsilon_{\text{s/w}}-i\frac{\sigma_{\text{s/w}}}{2\pi\nu}\approx \epsilon_{\text{s/w}}.
\eeq
In this expression, $\epsilon_{\text{s/w}}$ is the dielectric permittivity and $\sigma_{\text{s/w}}$ the conductance of silica (water).

\subsubsection{Polarizability of the metallic hemisphere}
For the metal hemisphere, the dominant polarization mechanism is the capacitance $C_\text{DL}=\epsilon_\text{w}/\lambda_D$ of the electrical double layer at the metal-water interface \cite{garcia-sanchez2012}, which is set by the Debye length $\lambda_D$. The Debye length for deionized water is approximately $\lambda_D=250$ nm \cite{yan2016}.

Without taking into account the silicon oxide layer, the Clausius-Mossotti function due to electric double layer capacitance is given by \cite{garcia-sanchez2012}
\beq
K_\text{m}(\nu)=\frac{(\nu/\nu_c)^2-1/2}{(\nu/\nu_c)^2+1}+i\frac{3}{2}\frac{(\nu/\nu_c)}{(\nu/\nu_c)^2+1},
\eeq
with the critical frequency $\nu_c=\frac{2\sigma}{RC_\text{DL}}$, where $\sigma$ is the conductivity of the deionized water.

However, the silicon oxide layer causes a shift in the critical frequency, $\nu_c\to\tilde\nu_c=(1+\delta_1)\nu_c$ with $\delta_1=\epsilon_w d_{\text{SiO}_2}/\lambda_D\epsilon_{\text{SiO}_2}$, as well as a renormalization $K(\nu)\to \tilde K(\nu)= K(\nu)/(1+\delta_2)$ with $\delta_2=\epsilon_w d_{\text{SiO}_2}/2R\epsilon_{\text{SiO}_2}$ \cite{pascall2010}.

It is important to emphasize that this calculation only gives a rough estimate for the metal-hemisphere dipole strength. Our calculation supposes that the metal coating is a perfect conductor, but the Titanium coatings in our experiments are so thin ($d_\text{Ti}=6$~nm) that they cease to be continuous. Thus, charges cannot redistribute freely across the metal hemisphere surface, which will affect both the dipole magnitude and position \cite{yan2015}. We therefore do not claim that this is quantitatively accurate, but it provides a reasonable estimate.

\subsection{Position of dipoles and dimer geometry}
In this section, we discuss the position of the electric dipoles in the Janus colloids. Since the colloid dimers form by two monomers aligning their metal dipoles, the dipole spacing also determines the overall geometry of the dimers.

\subsubsection{Position of the dipole in the dielectric hemisphere}
We assume that the electric dipole of the silica sphere is located at the center of mass of the induced surface charge distribution. Under this assumption, the position of this dipole is shifted by $\ell_\text{d}=3R/8$ from the center of the colloid, as calculated in Ref.~\cite{zhang2016}.

\subsubsection{Position of the dipole in the metallic hemisphere} \label{sisubsec:metaldipole}
The titanium coatings used in our experiments are so thin ($d_\text{Ti}=6$~nm) that the metal film is not continuous but rather patchy \cite{yan2015}. Hence, the effective conducting surface area is smaller and charges cannot redistribute freely between disconnected regions. As a result, the effective dipole shift from the center of the colloid will be larger than for the dielectric hemisphere. It is however difficult to obtain a theoretical estimate.

We therefore turn to experimental measurements of the effective dipole shift in magnetic Janus particles (Figure 1d in Ref.~\cite{yan2015}). By extrapolating this data to $d_\text{Ti}=6$~nm, we obtain an effective dipole shift of $\ell_\text{m}\approx0.56R$.

\begin{figure}[t!]
	\centering
	\includegraphics[width=0.4\textwidth]{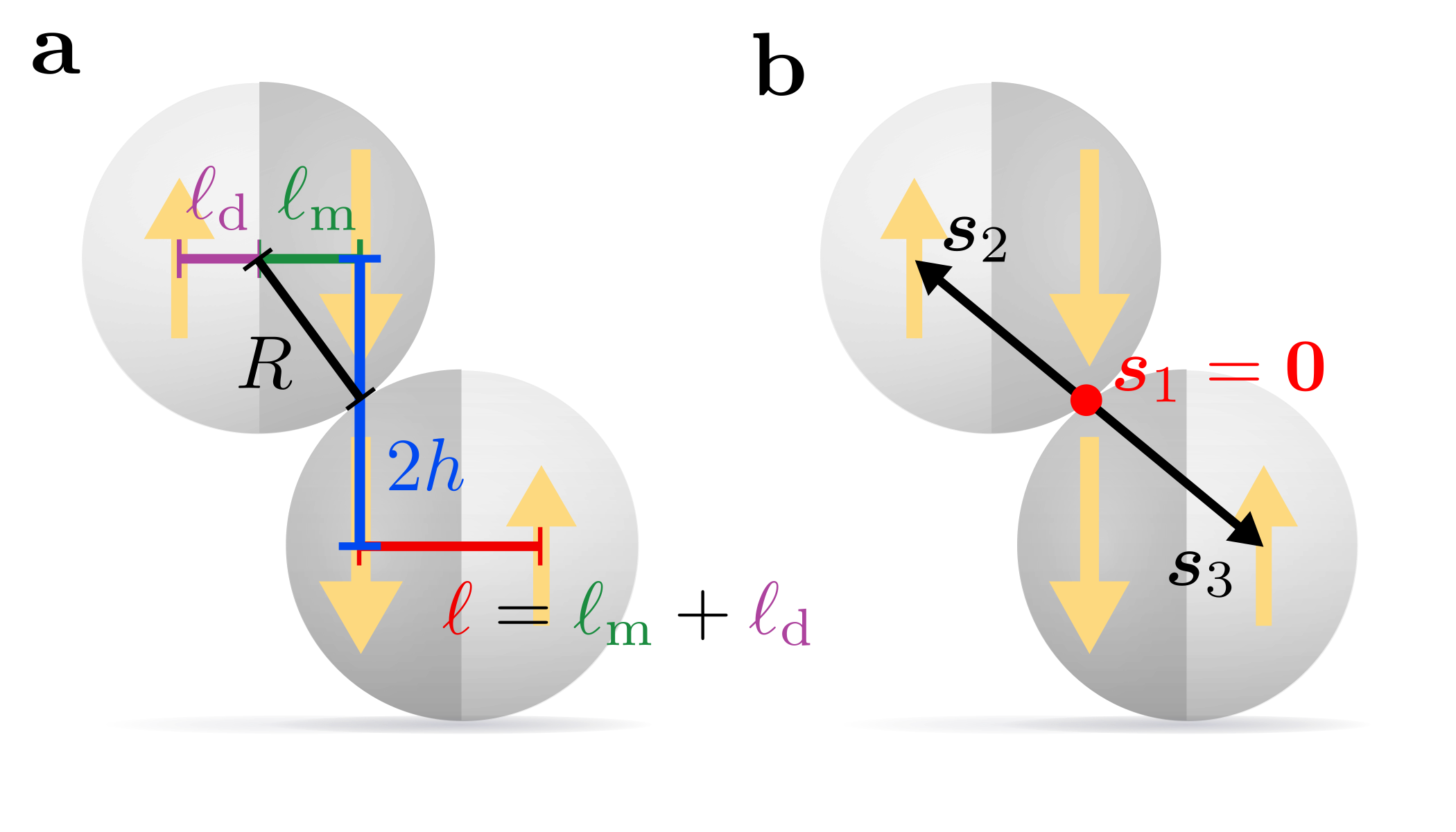}
    {\phantomsubcaption\label{sifig:dimer_geometry-lengths}}
    {\phantomsubcaption\label{sifig:dimer_geometry-vectors}}
	\caption{
		\textbf{Geometry of Janus colloid dimers.}
		Dimers consist of two monomers of radius $R$ whose metal hemispheres are touching such that the two induced dipoles in them are aligned. (\subref*{sifig:dimer_geometry-lengths}) The metal (dielectric) hemisphere dipoles are shifted from the center of mass of the monomers by a distance $\ell_\text{m}$ ($\ell_\text{d}$), such that the total distance between them is $\ell=\ell_\text{m}+\ell_\text{d}$. The vertical offset between the two monomers is $2h$. (\subref*{sifig:dimer_geometry-vectors}) In our model, each dimer $i$ contains three electric dipoles, shifted by the vectors $\vec s_{m_i}$, $m_i=1,2,3$ from the center of mass (red point).
	}  
	\label{sifig:dimer_geometry}
\end{figure}

As a result, the total distance between the metal and dielectric dipoles is $\ell=\ell_\text{m}+\ell_\text{d}\approx 0.935R\approx R$. We can also use the metal dipole shift $\ell_\text{m}$ to determine the vertical offset $2h$ between the head and tail monomers that form a dipole, as shown in \cref{sifig:dimer_geometry-lengths}, which gives $h=\sqrt{R^2-\ell_\text{m}^2}\approx 0.83R$. 

As in the case of the metal hemisphere dipole magnitude, we emphasize that $\ell_\text{m}$ (and the derived quantities $\ell$ and $h$) should only be considered approximate, since the irregular Titanium coatings of the colloids make an exact calculation impossible.

\subsection{Friction and diffusion coefficients of colloidal dimers}
The Janus particle monomers are spherical, so their translational and rotational friction coefficients $\xi_\text{t}^{\text{mono}}=6\pi\eta R$ and $\xi_\text{r}^{\text{mono}}=8\pi\eta R^3$ can be determined using standard Stokes drag. Here, $\eta=0.89$ mPa s is the viscosity of the surrounding fluid (water) and $R=1.5$ $\mu$m is the radius of the colloid. Dimers will experience stronger drag since they consist of two monomers, which in turn means their diffusion constants will be lower. 

\subsubsection{Friction coefficients}
To determine the translational and rotational friction coefficients of the dimer, we simplify the geometry of the dimer to consist of two spheres of radius $R$ lying in-plane and touching each other. The centers of the spheres are then located at $\vec r_1=R\vec x$ and $\vec r_2=-R\vec x$. Each colloid acts as a Stokeslet, and the resulting velocity field is a linear superposition of the individual flow fields. 

We can thus relate a force $\vec F_j$ acting on colloid $j$ to the velocity $\vec v_i$ it induces on colloid $i$ as \cite{kim1991}
\beq
\vec v_i=\tens{M}_{ij}\vec F_j.
\eeq
Stokes drag sets the self-mobility tensor to
\beq
\tens{M}_{11}=\tens{M}_{22}=\frac{1}{6\pi\eta R}\tens{I},
\eeq
proportional to the identity tensor $\tens{I}$, whereas for the influence of one colloid on the other, one solves the Stokes equation to obtain the mobility tensor
\beq
\tens{M}_{12}=\frac{1}{8\pi\eta}\tens G(\vec r_1-\vec r_2),
\eeq
where $\tens G(\vec r)$ is the Oseen tensor
\beq
\tens G(\vec r)=\frac{1}{r}\tens{I}+\frac{1}{r^3}\vec r\vec r.
\eeq
In our case, with $\vec r_{1}-\vec r_2= 2R\vec x$, the off-diagonal mobility tensors read
\beq
\tens{M}_{12}=\tens{M}_{21}=\frac{1}{16\pi\eta R}(\tens{I}+\vec x\vec x).
\eeq

Let us now consider a translation of the dimer along $\vec x$. In this case, $\vec v_1=\vec v_2=V\vec x$ and $\vec F_1=\vec F_2=F\vec x$, with a total force $\vec F_\text{tot}=2F\vec x$. Then,
\beq
\vec v_1=\tens{M}_{11}\vec F_1+\tens{M}_{12}\vec F_2=\(\frac{1}{6\pi\eta R}+\frac{1}{8\pi\eta R}\)\vec F,
\eeq
from which we deduce
\beq
\xi_\text{t}^\parallel=\frac{F_\text{tot}}{V}=\frac{2F}{V}=\frac{48\pi\eta R}{7}\approx 1.14\cdot\xi_\text{t}^\text{mono}.
\eeq
Similarly, we can consider a translation along $\vec y$ to compute $\xi_\text{t}^\perp=96\pi\eta R/11\approx 1.45\cdot\xi_\text{t}^\text{mono}$. To avoid the complexities of anisotropic friction, we use the mean of these two values in our simulations, $\xi_\text{t}^\text{dimer}=1.3\cdot\xi_\text{t}^\text{mono}$.

In determining the rotational friction coefficient, we need to take into account both the rotation of the two monomers around their own axis, which contributes $\xi_\text{r}^\text{mono}$ per monomer, and the translation of the monomer around the center of mass. The monomers move with velocities $\vec v_1=\Omega R\vec y$ and $\vec v_2=-\Omega R\vec y$, being subjected to forces $\vec F_1=F\vec y$ and $\vec F_2=-F\vec y$. Then,
\beq
\vec v_1=\Omega R\vec y=\tens{M}_{11}\vec F_1+\tens{M}_{12}\vec F_2=\(\frac{1}{6\pi\eta R}-\frac{1}{8\pi\eta R}\)\vec F_1,
\eeq
while the torque felt by the dimer is $\tau_z=R\cdot F+(-R)(-F)=2RF$. The friction coefficient for rotation around $\vec z$ due to translational movement of the spheres is then
\beq
\xi_\text{r}^{\text{tr},z}=\frac{\tau_z}{\Omega}=\frac{2R^2}{1/(6\pi\eta R)-1/(8\pi\eta R)}=\frac{95\pi\eta R^3}{5}.
\eeq
Together, we obtain the total friction coefficient due to both spinning and translation of the monomers:
\beq
\xi_\text{r}^\text{dimer}=2\xi_\text{r}^\text{mono}+\xi_\text{r}^{\text{tr},z}=\frac{176\pi\eta R^3}{5}\approx 4.4\cdot\xi_\text{r}^\text{mono}.
\eeq

\subsubsection{Diffusion coefficients}
As previous studies of this experimental system have shown, the rotational diffusion of the monomeric colloids is mainly caused by imperfections of the electrode surface \cite{zhang2021}. Due to this non-thermal origin of rotational fluctuations, we cannot use the rotational Stokes-Einstein relation $D_\text{r}^\text{SE}=k_BT/\xi_\text{r}$ to determine the effect of the increased rotational friction coefficient $\xi_\text{r}^\text{dimer}$ on the rotational diffusion coefficient $D_\text{r}^\text{dimer}$. However, we do not have experimental observations of single dimers that we could use to measure the rotational diffusion of a dimer directly.

Instead, let us consider fluctuations of the particle orientation $\theta$,
\beq
\xi_\text{r}\dot\theta=\tau(t),\label{eq:theta_flucts}
\eeq
where $\tau(t)$ is a random torque due to the inhomogeneous substrate.
By solving \cref{eq:theta_flucts}, we obtain the variance of $\theta$,
\beq
\sigma_\theta^2=\langle\theta^2(t)\rangle-\langle\theta(t)\rangle^2=\frac{1}{\xi_\text{r}^2}\int_0^t \dd t'\int_0^t \dd t''\langle\tau(t')\tau(t'')\rangle.
\eeq
We assume that the correlations of the random torque $\langle\tau(t')\tau(t'')\rangle$ are set by substrate properties and thereby independent of $\xi_\text{r}$. Thus $D_\text{r}\sim 1/\xi_\text{r}^2$, which implies that the rotational diffusion coefficient of a dimer is $4.4^2\approx 20$ times lower than that of a monomer.

Translational diffusion is already irrelevantly small for the monomeric Janus colloids \cite{zhang2021}, and even lower for the larger dimers. We therefore neglect it in 
our simulations.

\section{Microscopic model}\label{si:model}
Here, we introduce the microscopic model used in our simulations in detail. As stated in the Main Text, we treat each dimer $i$ as a single Active Brownian Particle with center of mass $\vec r_i$ that propels with velocity $v_0$ along $\hat{\vec n}_i(\theta_i)=(\cos\theta_i,\sin\theta_i)^T$, where $\theta_i$ is the particle orientation. Particles interact via pairwise forces $\vec F_{ji}$ and torques $\Gamma_{ji}$. We consider only the electrostatic interactions between the induced electric dipoles. In particular, we neglect steric and hydrodynamic interactions. Additionally, particles are subject to rotational noise $D_\text{r}$, as well as translational and rotational friction coefficients $\xi_\text{t}$ and $\xi_\text{r}$. This is described by the Langevin equations
\begin{subequations} \label{eq:eom_si}
\begin{align}
    \frac{\dd\vec r_i}{\dd t}=&\,v_0\hat{\vec n}_i(\theta_i)+\frac{1}{\xi_\text{t}}\sum_{j\neq i}\vec F_{ji}, \label{eq:eom_f_si}\\
    \frac{\dd \theta_i}{\dd t}=&\,\frac{1}{\xi_\text{r}}\sum_{j\neq i} \Gamma_{ji}+\sqrt{2D_\text{r}}\eta_i(t), \label{eq:eom_t_si}
\end{align}
\end{subequations}
where $\eta_i(t)$ is a Gaussian white noise that satisfies $\langle \eta_i(t)\eta_j(t')\rangle=\delta_{ij}\delta(t-t')$.

To compute the force and torque between two particles, we must add up the contributions from all pairwise interactions between the electric dipoles of the particles. Since each monomer has two induced dipoles, one in the dielectric (with magnitude $\vec d_\text{d}$) and another in the metal hemisphere (with magnitude $\vec d_\text{m}$), dimer $i$ contains four dipoles which we label with index $m_i$. The interaction force $\vec F_{ji}$ and torque $\vec \Gamma_{ji}$ felt by a dimer at position $\vec r_i$ with orientation $\theta_i$ due to a dimer at position $\vec r_j$ with orientation $\theta_j$ are then:
\begin{subequations}
\begin{align}
\vec F_{ji}&=e^{-|\vec r_i-\vec r_j|/\lambda}\sum_{m_i,m_j} \vec F(m_i,m_j) \label{eq:pairwise_F}\\
\vec \Gamma_{ji} &=e^{-|\vec r_i-\vec r_j|/\lambda}\sum_{m_i,m_j} \vec s_{m_i}\times \vec F(m_i,m_j). \label{eq:pairwise_T}
\end{align}
\end{subequations}
Here, $\vec F(m_i,m_j)=-\vec\nabla U(m_i,m_j)$ is the force between dipoles $m_i$ and $m_j$, and $\vec s_{m_i}$ is the offset of dipole $m_i$ from the center of mass of dimer $i$ (see \cref{sifig:dimer_geometry-vectors}). We also account for the fact that the electrodes, separated by $\lambda=120$ $\mu$m, screen the interactions, by adding an exponential prefactor. The energy $U$ of two dipoles is
\beq
U(m_i,m_j)=\frac{\vec d_{m_i}\cdot \vec d_{m_j}-3(\vec d_{m_i}\cdot \hat{\vec r}_{m_i,m_j})(\vec d_{m_j}\cdot \hat{\vec r}_{m_i,m_j})}{4\pi\epsilon_\text{w}|\vec r_{m_i,m_j}|^3},
\eeq
where $\epsilon_\text{w}$ is the dielectric permittivity of the solvent, $\vec r_{m_i,m_j}=(\vec r_i+\vec s_{m_i})-(\vec r_j+\vec s_{m_j})$ is the vector pointing from one dipole to the other. The corresponding unit vector is $\hat{\vec r}_{m_i,m_j}=\vec r_{m_i,m_j}/|\vec r_{m_i,m_j}|$.

In the experiments, particles can only rotate around the $z$ axis because the electric field keeps the particle equator oriented perpendicular to the field. This is likely due to dielectrophoretic effects, although induced-charge electro-osmosis may also play a role \cite{gangwal2008}. We therefore only consider torques along the $z$-axis, $\Gamma_{ji}=\hat{\vec z}\cdot\vec\Gamma_{ji}$.

One simplification proves helpful. Computing $4^2=16$ pairwise interactions is computationally heavy, and we do not know how charges redistribute between the two metal-coated hemispheres, in particular due to their irregular structure (see \cref{sisubsec:metaldipole}). Therefore, we replace the two metal-hemisphere dipoles by a single dipole of twice the magnitude, $2\vec d_\text{m}$, sitting in the center of mass of the dimer, leaving only $3^2=9$ interactions to compute.

To summarize, a dimer $i$ in our model contains three electric dipoles $m_i=1,2,3$, with magnitudes $\vec d_{m_i}$ and positions $\vec r_i+\vec s_{m_i}$: The first is the central dipole of magnitude $\vec d_1=2\vec d_\text{m}$ located at the center of mass, i.e. $\vec s_1=\vec 0$. The other two are the outer electric dipoles of magnitude $\vec d_2=\vec d_3=\vec d_\text{d}$ which are shifted away from the center of mass by $\vec s_2=-\vec s_3=\ell\hat{\vec n}_i(\theta_i)+h\hat{\vec z}$. We use \cref{eq:pairwise_F,eq:pairwise_T} to compute the interactions between these particles.

\section{Coarse-graining}\label{si:coarsegrain}
In this section, we derive the hydrodynamic theory describing the collective behavior of the Janus particle dimers based on the microscopic equations of motion \cref{eq:eom_si}. We do this by extending the methods used in Refs.~\cite{zhang2021,das2024}: First, we derive the dynamics of the one-particle distribution function $\Psi_1(\vec r_1,\theta_1;t)$ from the Smoluchowski equation, which describes the evolution of the entire $N$-particle system using the $N$-particle distribution function $\Psi_N(\vec r_1,\theta_1,\vec r_2,\theta_2,...,\vec r_N,\theta_N;t)$.

We then expand this equation using the Bogoliubov?Born?Green?Kirkwood?Yvon (BBGKY) hierarchy, which we truncate at the level of pair correlations. We make the interaction terms local by expanding them in gradients and angular Fourier modes. After this, we define continuum fields like the density field $\rho(\vec r,t)$, the polarization density field $p_\alpha(\vec r,t)$ or the nematic order density field $Q_{\alpha\beta}(\vec r,t)$, as the moments of the one-particle distribution function $\Psi_1(\vec r_1,\theta_1;t)$.

As a final step, we use symmetry arguments --- in particular the chiral symmetry of our system --- to simplify the hydrodynamic equations, arriving at the final coarse-grained theory.

\subsection{Smoluchowski equation}
Rather than describing our system with the Langevin equations \cref{eq:eom_si}, we can equivalently use the corresponding Fokker-Planck equation for the full $N$-particle distribution function $\Psi_N(\vec r_1,\theta_1,\vec r_2,\theta_2,...,\vec r_N,\theta_N;t)$, which is also called the Smoluchowski equation:
\beq
\partial_t \Psi_N=-\sum_{i=1}^N \( \vec\nabla_i\cdot \vec J_{\text{t},i}+\ni\times\partial_{\ni}\cdot \vec  J_{\text{r},i} \).\label{eq:smoluchowskiN}
\eeq
In this equation, $\vec\nabla_i$ is the gradient operator with respect to the position of particle $i$. Similarly, $\vec n_i=(\cos\theta_i,\sin\theta_i)^T$ is the orientation and $\ni\times\partial_{\ni}$ is the rotation operator for particle $i$. The translational probability current $\vec J_{\text{t},i}$ and the rotational probability current $\vec J_{\text{r},i}$ are given by
\begin{subequations}
\begin{align}
\vec J_{\text{t},i}&=\(v_0\ni+\frac{\vec F_i}{\xi_\text{t}}\)\Psi_N-D_\text{t}\vec\nabla_i\Psi_N, \\
\vec J_{\text{r},i}&=\frac{\vec\Gamma_i}{\xi_\text{r}}\Psi_N-D_\text{r}\ni\times\partial_{\ni}\Psi_N,
\end{align}
\end{subequations}
which contain drift terms due to particle self-propulsion and interactions as well as diffusion terms due to positional and rotational noise. We keep the spatial diffusion term $D_\text{t}$ in our calculations for the sake of completeness, even though we neglect it in our simulations.

By integrating out the degrees of freedom of $N-1$ particles, we obtain the evolution equation of the one-particle distribution function $\Psi_1(\vec r_1,\theta_1;t)$, which reads
\begin{align}
\partial_t\Psi_1(\vec r_1,\theta_1;t)=&-\vec\nabla_1\cdot\[(v_0\hat{\vec n}_1-D_\text{t}\vec\nabla_1)\Psi_1\]+D_\text{r}\partial_{\theta_1}^2\Psi_1 \nonumber\\&-\vec\nabla_1\cdot\frac{\vec F^\text{int}}{\xi_\text{t}}-\partial_{\theta_1}\frac{\hat{\vec z}\cdot\vec\Gamma^\text{int}}{\xi_\text{r}}.\label{eq:smoluchowski1}
\end{align}
In this expression, all interaction effects are contained in the so-called collective force and torque,
\begin{subequations}
\begin{align}
\vec F^\text{int}&=\int \dd\vec r_2 \dd\theta_2 \vec F(\vec r_1,\theta_1,\vec r_2,\theta_2)\Psi_2(\vec r_1,\theta_1,\vec r_2,\theta_2;t),\label{eq:fint}\\
\vec \Gamma^\text{int}&=\int \dd\vec r_2 \dd\theta_2 \vec\Gamma(\vec r_1,\theta_1,\vec r_2,\theta_2)\Psi_2(\vec r_1,\theta_1,\vec r_2,\theta_2;t),\label{eq:gammaint}
\end{align}
\end{subequations}
where $\vec F(\vec r_1,\theta_1,\vec r_2,\theta_2)$ and $\vec \Gamma(\vec r_1,\theta_1,\vec r_2,\theta_2)$ are respectively the force and torque felt by a particle at position $\vec r_1$ with orientation $\theta_1$ due to a particle at position $\vec r_2$ with orientation $\theta_2$, which are given in \cref{si:model}.

The collective force and torque also depend on the two-particle distribution function $\Psi_2(\vec r_1,\theta_1,\vec r_2,\theta_2;t)$, which contains information about pair correlations between particles. Instead of determining $\Psi_2$ from its equation of motion, which will depend on $\Psi_3$, etc., we close the hierarchy by writing
\begin{align}
\Psi_2&(\vec r_1,\theta_1,\vec r_2,\theta_2;t) \nonumber \\ &=\Psi_1(\vec r_2,\theta_2;t)g(\vec r_2,\theta_2|\vec r_1,\theta_1;t)\Psi_1(\vec r_1,\theta_1;t),
\end{align}
where we have introduced the pair distribution function $g(\vec r_2,\theta_2|\vec r_1,\theta_1;t)$ that describes the conditional probability of finding a particle at position $\vec r_2$ with orientation $\theta_2$, given that another particle is located at $\vec r_1$ with orientation $\theta_1$. Frequently, the mean-field approximation $g\approx 1$ is used, but, as we shall see, the pair correlations between dimers play an important role in the formation of streams. Instead, we measure the pair distribution function numerically as described in the \hyperref[sec:methods]{Methods}.

Assuming that the system is in a homogeneous steady state, the pair distribution function will be independent of time and only depend on relative positions and orientations of particles due to translational invariance. We thus write
\beq
g(\vec r_2,\theta_2|\vec r_1,\theta_1;t)=g(|\vec r_2-\vec r_1|,\varphi,\theta_2-\theta_1),
\eeq
where we have defined the interparticle distance $r=|\vec r|=|\vec r_2-\vec r_1|$, the positional angle $\varphi$, defined by $\cos\varphi=\hat{\vec n}_1\cdot\hat{\vec r}$ (i.e. the angle between the interparticle vector $\vec r$ and the orientation vector $\hat{\vec n}_1$ of particle 1) as well as the relative orientation angle $\theta=\theta_2-\theta_1$ (see also Refs.~\cite{zhang2021,das2024}).

Similarly, we can express the pairwise force and torque appearing in the collective force and torque using these coordinates,
\begin{subequations} \label{eq:coll_si}
\begin{align}
\vec F(\vec r_1,\theta_1,\vec r_2,\theta_2)&=-F(r,\varphi,\theta)\hat{\vec r}+F_\perp(r,\varphi,\theta)\hat{\vec r}_\perp, \label{eq:collforce}\\
\vec\Gamma(\vec r_1,\theta_1,\vec r_2,\theta_2)&=\Gamma(r,\varphi,\theta)\vec z. \label{eq:colltorque}
\end{align}
\end{subequations}
We have here decomposed the force into a central force $F$ and a non-central component $F_\perp$, which arises due to the interacting electric dipoles not being located in the center of mass. In the following, we assume forces to be central, i.e. $F_\perp\approx 0$, justified by the fact that these non-central components of the force involve the weaker electric dipoles in the dielectric hemispheres of the colloids, and that these effects would only be relevant at very short interparticle distances.

Combining these steps, we express the collective force and torque as
\begin{subequations}
\begin{align}
\vec F^\text{int}&=-\Psi_1(\vec r_1,\theta_1;t)\times \nonumber \\ &\qquad\int \dd^2\vec r_2 \dd\theta_2\, F(r,\varphi,\theta)\Psi_1(\vec r_2,\theta_2)g(r,\varphi,\theta)\hat{\vec r}(\varphi),\\
\vec \Gamma^\text{int}&=\Psi_1(\vec r_1,\theta_1;t)\times \nonumber \\ &\qquad\int \dd^2\vec r_2 \dd\theta_2\, \Gamma(r,\varphi,\theta)\Psi_1(\vec r_2,\theta_2)g(r,\varphi,\theta)\hat{\vec z},
\end{align}
\end{subequations}
where we recall that $r=|\vec r|=|\vec r_2-\vec r_1|$, $\theta=\theta_2-\theta_1$ while $\varphi$ is the angle between $\hat{\vec n}_1$ and $\vec r$.

\subsection{Gradient and Fourier expansions of interactions}
The above expressions for the collective force and torque depend non-locally on the one-particle distribution $\Psi_1$. To obtain local hydrodynamic equations, we expand $\Psi_1$ both in spatial gradients and angular Fourier modes,
\begin{align}
\Psi_1(&\vec r_2,\theta_2)\approx \nonumber\\ & \frac{1}{2\pi}\sum_{k=-\infty}^\infty e^{-ik\theta_2}\(\tilde\Psi_{1,k}(\vec r_1) + \vec\nabla_1\tilde\Psi_{1,k}(\vec r_1)\cdot(\vec r_2 - \vec r_1) \) \nonumber \\ &\quad+\mathcal O(\vec\nabla_1^2).
\end{align}
Here, $\vec\nabla_1$ is the gradient operator with respect to the position of particle 1. For the sake of simplicity, we truncate the expansions at linear order in gradients and at order $k=\pm 1$ in Fourier modes.

By inserting this expansion into \cref{eq:coll_si}, we rewrite the collective force and torque as
\begin{subequations}
\begin{align}
\vec F^\text{int}&\approx -\sum_{k=-1}^1 e^{-ik\theta_1}(\vec F^{0,k}(\vec r_1,\theta_1)+\vec F^{1,k}(\vec r_1,\theta_1)), \\
\vec\Gamma^\text{int}&\approx \sum_{k=-1}^1 e^{-ik\theta_1}\(\Gamma^{0,k}(\vec r_1,\theta_1)+\Gamma^{1,k}(\vec r_1,\theta_1)\)\hat{\vec z}
\end{align}
\end{subequations}
with
\begin{widetext}
\begin{subequations}
\begin{align}
\vec F^{0,k}(\vec r_1,\theta_1)&=\Psi_1(\vec r_1,\theta_1)\(\int_0^\infty r\, \dd r\int_0^{2\pi}\dd\varphi\int_0^{2\pi}\dd\theta\, \hat{\vec r}(\varphi)F(r,\varphi,\theta)e^{-ik\theta}g(r,\varphi,\theta)\)\tilde\Psi_{1,k}(\vec r_1),\\
\vec F^{1,k}(\vec r_1,\theta_1)&=\Psi_1(\vec r_1,\theta_1)\(\int_0^\infty r^2\, \dd r\int_0^{2\pi}\dd\varphi\int_0^{2\pi}\dd\theta\, \hat{\vec r}(\varphi) F(r,\varphi,\theta)e^{-ik\theta}g(r,\varphi,\theta)(\hat {\vec r}(\varphi)\cdot\vec\nabla_1)\)\tilde\Psi_{1,k}(\vec r_1),\\
\Gamma^{0,k}(\vec r_1,\theta_1)&=\Psi_1(\vec r_1,\theta_1)\(\int_0^\infty r\, \dd r\int_0^{2\pi}\dd\varphi\int_0^{2\pi}\dd\theta\, \Gamma(r,\varphi,\theta)e^{-ik\theta}g(r,\varphi,\theta)\)\tilde\Psi_{1,k}(\vec r_1),\\
\Gamma^{1,k}(\vec r_1,\theta_1)&=\Psi_1(\vec r_1,\theta_1)\(\int_0^\infty r^2\, \dd r\int_0^{2\pi}\dd\varphi\int_0^{2\pi}\dd\theta\, \Gamma(r,\varphi,\theta)e^{-ik\theta}g(r,\varphi,\theta)(\hat {\vec r}(\varphi)\cdot\vec\nabla_1)\)\tilde\Psi_{1,k}(\vec r_1).
\end{align}
\end{subequations}
\end{widetext}
Here, the superscripts of $\vec F$ and $\Gamma$ refer to the orders in the gradient and angular Fourier expansions, respectively. Recall that the set of integration coordinates $r,\varphi,\theta$ are defined in a reference frame relative to particle 1, and that the components of the unit vector $\hat{\vec r}(\varphi)=(\cos\varphi,\sin\varphi)^T$ in the integrals above are in the basis formed by $\hat{\vec n}_1$ and $\hat{\vec z}\times\hat{\vec n}_1$. In the lab frame we have $\hat{\vec n}_1=(\cos\theta_1,\sin\theta_1)^T$, and hence
\beq
\hat{\vec r}(\varphi)=\tens{ R}(\varphi)\cdot \hat{\vec n}(\theta_1)=\begin{pmatrix}
	\cos\varphi & -\sin\varphi\\
	\sin\varphi & \cos\varphi\\
\end{pmatrix}\cdot \begin{pmatrix}
	\cos\theta_1\\
	\sin\theta_1\\
\end{pmatrix}.
\eeq
We can use this relation to write the collective force and torque in a succinct manner by identifying the integrals in the equations above as coefficients of an expansion of the interactions. Switching to Einstein notation for summation over repeated indices, we rewrite the collective force and torque as
\begin{subequations}
\begin{align}
F_\alpha^\text{int}&=-\Psi_1(\vec r_1,\theta_1) \,\times \nonumber \\ &\quad\sum_{k=-1}^1 e^{-ik\theta_1}\(\zeta_{\alpha\beta}^{0,k}n_\beta+ \zeta_{\alpha\beta\gamma\delta}^{1,k}n_\gamma n_\delta\partial_\beta\)\tilde\Psi_{1,k}(\vec r_1), \\
\Gamma^\text{int}&=\Psi_1(\vec r_1,\theta_1)\sum_{k=-1}^1 e^{-ik\theta_1}\(\tau^{0,k}+ \tau_{\alpha\beta}^{1,k}n_\beta\partial_\alpha\)\tilde\Psi_{1,k}(\vec r_1),
\end{align}
\end{subequations}
where $n_\alpha$ is the component $\alpha\in\{x,y\}$ of the orientation vector $\hat{\vec n}_1=(\cos\theta_1,\sin\theta_1)^T$, and the expansion coefficients read
\begin{widetext}
\begin{subequations} \label{eq:zetatau}
\begin{align}
\zeta^{0,k}_{\alpha\beta}&= \frac{1}{2\pi}\int_0^\infty r\, \dd r\int_0^{2\pi}\dd\varphi\int_0^{2\pi}\dd\theta\, F(r,\varphi,\theta)e^{-ik\theta}g(r,\varphi,\theta) R_{\alpha\beta}(\varphi), \label{eq:zeta0k}\\
\zeta_{\alpha\beta\gamma\delta}^{1,k}&= \frac{1}{2\pi}\int_0^\infty r^2\, \dd r\int_0^{2\pi}\dd\varphi\int_0^{2\pi}\dd\theta\, F(r,\varphi,\theta) e^{-ik\theta}g(r,\varphi,\theta) R_{\alpha\gamma}(\varphi) R_{\beta\delta}(\varphi), \label{eq:zeta1k}\\
\tau^{0,k}&=\frac{1}{2\pi}\int_0^\infty r\, \dd r\int_0^{2\pi}\dd\varphi\int_0^{2\pi}\dd\theta\, \Gamma(r,\varphi,\theta)e^{-ik\theta}g(r,\varphi,\theta), \label{eq:tau0k}\\
\tau_{\alpha\beta}^{1,k}&= \frac{1}{2\pi}\int_0^\infty r^2\, \dd r\int_0^{2\pi}\dd\varphi\int_0^{2\pi}\dd\theta\, \Gamma(r,\varphi,\theta) e^{-ik\theta}g(r,\varphi,\theta) R_{\alpha\beta}(\varphi). \label{eq:tau1k}
\end{align}
\end{subequations}
\end{widetext}
All information about the interactions is now contained in these coefficients. We note that $(\zeta^{0,k}_{\alpha\beta})^*=\zeta^{0,-k}_{\alpha\beta}$, and similarly for the other coefficients.

\subsection{Derivation of hydrodynamic equations}
We can use the expressions derived in the previous sections to obtain hydrodynamic equations for fields like the particle density $\rho(\vec r,t)$, polarization density $p_\alpha(\vec r,t)$ or the nematic order density $\tens Q(\vec r,t)$ by observing that these fields are the angular moments of the one-particle distribution function $\Psi_1$:
\begin{subequations}
\begin{align}
\rho(\vec r)&=\int \dd\theta\, \Psi_1(\vec r,\theta),\\
p_\alpha(\vec r)&=\int \dd\theta\, n_\alpha(\theta)\Psi_1(\vec r,\theta),\\
Q_{\alpha\beta}(\vec r)&=\int \dd\theta\, \(n_\alpha n_\beta-\frac{1}{2}\delta_{\alpha\beta}\)\Psi_1(\vec r,\theta).
\end{align}
\end{subequations}
Higher-order moments are similarly defined to be traceless:
\begin{subequations}
\begin{align}
&  T_{\alpha\beta\gamma}(\vec r,t) \nonumber\\&=\int \dd\theta\,\(n_\alpha n_\beta n_\gamma-\frac{1}{4}(\delta_{\alpha\beta}n_\gamma+\delta_{\alpha\gamma}n_\beta+\delta_{\beta\gamma}n_\alpha)\)\Psi_1(\vec r,\theta),\\
&  V_{\alpha\beta\gamma\delta}(\vec r,t)\nonumber\\ &=\int \dd\theta\,\bigg(n_\alpha n_\beta n_\gamma n_\delta-\frac{1}{6}(\delta_{\alpha\beta}n_{\gamma}n_{\delta}+\text{permutations})
\nonumber \\ &\hspace{2cm} +\frac{1}{24}(\delta_{\alpha\beta}\delta_{\gamma\delta}+\delta_{\alpha\gamma}\delta_{\beta\delta}+\delta_{\alpha\delta}\delta_{\beta\gamma})\bigg)\Psi_1(\vec r,\theta).
\end{align}
\end{subequations}
We will neglect moments beyond the rank 2 (nematic) tensor, yet these definitions are relevant since lower-order traceless tensors like $p$ or $\tens Q$ will appear when splitting a high-rank tensor into its trace and traceless parts, contributing important terms in the hydrodynamic equations.

These moments can also be expressed in terms of the Fourier modes $\tilde\Psi_{1,k}$, in particular
\begin{subequations}
\begin{align}
\rho(\vec r,t)&=\tilde\Psi_{1,0}(\vec r,t),\\
\vec p(\vec r,t)&=
\begin{pmatrix}
	\text{Re}\,\tilde\Psi_{1,1}(\vec r,t),\\
	\text{Im}\,\tilde\Psi_{1,1}(\vec r,t)\\
\end{pmatrix}, \\
\tens Q(\vec r,t)&=\frac{1}{2}
\begin{pmatrix}
	\text{Re}\,\tilde\Psi_{1,2}(\vec r,t) & \text{Im}\,\tilde\Psi_{1,2}(\vec r,t)\\
	\text{Im}\,\tilde\Psi_{1,2}(\vec r,t) & -\text{Re}\,\tilde\Psi_{1,2}(\vec r,t)\\
\end{pmatrix}
\end{align}
\end{subequations}

We can now obtain hydrodynamic equations of motion for these fields by projecting \cref{eq:smoluchowski1} onto the corresponding moment of $\Psi_1$. Let us briefly recall this equation here and rewrite it in Einstein notation:
\begin{align}
\partial_t\Psi_1(\vec r,\theta;t)=&-\partial_\alpha\[(v_0 n_\alpha-D_\text{t}\partial_\alpha)\Psi_1\]+D_\text{r}\partial_\theta^2\Psi_1 \nonumber\\& -\frac{1}{\xi_\text{t}}\partial_\alpha F^\text{int}_{\alpha}-\frac{1}{\xi_\text{r}}\partial_\theta\Gamma^\text{int}.\label{eq:smoluchoswki1e}
\end{align}

To obtain the dynamics of the density field, we integrate out the angle $\theta$ from $\Psi_1$:
\begin{align}
\partial_t\rho(\vec r,t)&=\partial_t\int_0^{2\pi} \dd\theta\, \Psi_1(\vec r,\theta;t)=\int_0^{2\pi} \dd\theta\, \partial_t\Psi_1(\vec r,\theta;t) \nonumber\\ 
&=-\partial_\alpha\((v_0 p_\alpha-D_\text{t}\partial_\alpha\rho)+\frac{1}{\xi_\text{t}}\int_0^{2\pi} \dd\theta\,F_\alpha^{\text{int}}\),
\end{align}
where we have identified the polarity as the integral in the self-propulsion term $v_0$ and the density in the diffusion term $D_\text{t}$.

Next, we evaluate the integral of the collective force. We omit the time dependence of the fields and integration bounds for $\theta$, integrated from $0$ to $2\pi$, to lighten the notation:
\begin{widetext}
\begin{align}
\int \dd\theta\, F_\alpha^\text{int}
&=-\int \dd\theta\,\Psi_1(\vec r,\theta)\big(\zeta_{\alpha\beta}^{0,0}n_\beta+\zeta_{\alpha\beta\gamma\delta}^{1,0}n_\gamma n_\delta \partial_\beta\big)\tilde\Psi_{1,0}(\vec r) -2\int \dd\theta\,\Psi_1(\vec r,\theta)\,\text{Re}\(e^{-i\theta}\big(\zeta_{\alpha\beta}^{0,1}n_\beta+\zeta_{\alpha\beta\gamma\delta}^{1,1}n_\gamma n_\delta \partial_\beta\big)\tilde\Psi_{1,1}(\vec r)\) \nonumber\\ 
&=-\zeta^{0,0}_{\alpha\beta}\rho(\vec r) p_\beta(\vec r)-\zeta^{1,0}_{\alpha\beta\gamma\delta}\big(\partial_\beta\rho(\vec r)\big)\(Q_{\gamma\delta}(\vec r)+\frac{1}{2}\delta_{\gamma\delta}\rho(\vec r)\) \nonumber\\
&\quad-2\Re (\zeta^{0,1}_{\alpha\beta})\Re\(\int \dd\theta \Psi_1(\vec r,\theta) e^{-i\theta} n_\beta \tilde\Psi_{1,1}(\vec r)\) +2\Im (\zeta^{0,1}_{\alpha\beta})\Im\(\int \dd\theta \Psi_1(\vec r,\theta) e^{-i\theta} n_\beta \tilde\Psi_{1,1}(\vec r)\) \nonumber\\
&\quad-2\Re (\zeta^{1,1}_{\alpha\beta\gamma\delta})\Re\(\int \dd\theta \Psi_1(\vec r,\theta) e^{-i\theta} n_\gamma n_\delta \partial_\beta \tilde\Psi_{1,1}(\vec r)\) +2\Im (\zeta^{1,1}_{\alpha\beta\gamma\delta})\Im\(\int \dd\theta \Psi_1(\vec r,\theta) e^{-i\theta} n_\gamma n_\delta \partial_\beta \tilde\Psi_{1,1}(\vec r)\).
\end{align}

The different terms can now be evaluated using the identity $\cos\theta\, p_x+\sin\theta\, p_y=n_\gamma p_\gamma$, for example as
\begin{align}
\Re & \(\int  \dd\theta \Psi_1(\vec r,\theta) e^{-i\theta} n_\beta \tilde\Psi_{1,1}(\vec r)\) =\Re\(\int \dd\theta \Psi_1(\vec r,\theta) (\cos\theta-i\sin\theta) n_\beta (p_x+ip_y)\) \nonumber\\ 
&=\int \dd\theta\,\Psi_1(\vec r,\theta) n_\beta (\cos\theta p_x+\sin\theta p_y) =\(\int \dd\theta\, \Psi_1(\vec r,\theta) n_\beta n_\gamma \)p_\gamma  =\(Q_{\beta\gamma}+\frac{1}{2}\delta_{\beta\gamma}\rho\)p_\gamma.
\end{align}

Having performed these integrals, we write out the dynamics of the density field. Since the particle number is a conserved quantity, it takes the form of a continuity equation $\partial_t\rho= -\partial_\alpha J_\alpha$, where $J_\alpha$ is the density current. The calculation for polarization and nematic order density fields follows the same principles. We obtain the following expressions:
\begin{subequations} \label{eq:hydrodynamic_eq_general}
\begin{align}
\partial_t\rho=&-\partial_\alpha J_\alpha, \label{eq:densdynfull}\\
J_\alpha =& \,v_0p_\alpha-D_\text{t}\partial_\alpha\rho-\frac{\zeta_{\alpha\beta}^{0,0}}{\xi_\text{t}}\rho p_\beta-\frac{\zeta_{\alpha\beta\gamma\delta}^{1,0}}{\xi_\text{t}}(\partial_\beta\rho)\(Q_{\gamma\delta}+\frac{1}{2}\delta_{\gamma\delta}\rho\) \nonumber \\
&-\frac{2\Re(\zeta_{\alpha\beta}^{0,1})}{\xi_\text{t}}\(Q_{\beta\gamma}+\frac{1}{2}\delta_{\beta\gamma}\rho\)p_\gamma +\frac{2\Im(\zeta_{\alpha\beta}^{0,1})}{\xi_\text{t}}\epsilon_{\gamma\delta}\(Q_{\beta\delta}+\frac{1}{2}\delta_{\beta\delta}\rho\)p_\gamma \nonumber\\
&-\frac{2\Re(\zeta_{\alpha\beta\gamma\delta}^{1,1})}{\xi_\text{t}}\( T_{\mu\gamma\delta}\partial_\beta p_\mu+\frac{1}{4}\(p_\delta \partial_\beta p_\gamma + p_\gamma \partial_\beta p_\delta + \delta_{\gamma\delta} p_\mu \partial_\beta p_\mu \)\) \nonumber \\
&+\frac{2\Im(\zeta_{\alpha\beta\gamma\delta}^{1,1})\epsilon_{\mu\nu}}{\xi_\text{t}}\( T_{\gamma\delta\nu}+\frac{1}{4}\(\delta_{\gamma\delta}p_\nu + \delta_{\gamma\nu}p_\delta + \delta_{\delta\nu}p_\mu \)\)\partial_\beta p_\mu, \label{eq:denscurrentfull}
\end{align}
\begin{align}
\partial_t p_\alpha=&-D_\text{r} p_\alpha - \frac{v_0}{2}\partial_\alpha\rho+D_\text{t}\partial_\beta\partial_\beta p_\alpha+\frac{\zeta_{\beta\alpha}^{0,0}}{2\xi_\text{t}}\partial_\beta(\rho^2)+\frac{\zeta_{\mu\beta\gamma\delta}^{1,0}}{2\xi_\text{t}}\partial_\mu\big( (\delta_{\alpha\gamma}p_\delta + \delta_{\alpha\delta}p_\gamma + \delta_{\gamma\delta}p_\alpha )\partial_\beta\rho \big) \nonumber \\
&+\frac{\Re( \zeta_{\mu\beta}^{0,1})}{\xi_\text{t}}\partial_\mu\(p_\alpha p_\beta +\frac{1}{2}\delta_{\alpha\beta} p^2\)-\frac{\Im (\zeta_{\mu\beta}^{0,1})}{2\xi_\text{t}}\partial_\mu(\epsilon_{\rho\alpha}p_\beta p_\rho + \epsilon_{\rho\beta}p_\alpha p_\rho) \nonumber \\
&+\frac{\Re( \zeta_{\mu\beta\gamma\delta}^{1,1})}{4\xi_\text{t}}\partial_\mu\big( (\delta_{\gamma\delta}\partial_\beta p_\alpha + \delta_{\alpha\delta}\partial_\beta p_\gamma + \delta_{\alpha\gamma}\partial_\beta p_\delta )\rho \big) -\frac{\Im (\zeta_{\mu\beta\gamma\delta}^{1,1})}{4\xi_\text{t}}(\delta_{\alpha\gamma}\epsilon_{\rho\delta}+\delta_{\alpha\delta}\epsilon_{\rho\gamma}+\delta_{\gamma\delta}\epsilon_{\rho\alpha})\rho\partial_\beta p_\rho \nonumber \\
&+\frac{\tau^{0,0}+\Re(\tau^{0,1})}{\xi_\text{r}}\epsilon_{\alpha\beta}\rho p_\beta -\frac{\Im(\tau^{0,1})}{\xi_\text{r}}\rho p_\alpha + \frac{\epsilon_{\alpha\beta}\tau^{1,0}_{\mu\beta}}{2\xi_\text{r}}\rho\partial_\mu\rho+\frac{\epsilon_{\alpha\beta}\Re(\tau^{1,1}_{\delta\gamma})}{2\xi_\text{r}}(\delta_{\beta\gamma}p_\mu + \delta_{\beta\mu}p_\gamma + \delta_{\gamma\mu}p_\beta)\partial_\delta p_\mu \nonumber \\
&-\frac{\epsilon_{\alpha\beta}\epsilon_{\mu\nu}\Im(\tau^{1,1}_{\delta\gamma})}{2\xi_\text{r}}(\delta_{\beta\gamma}p_\nu + \delta_{\beta\nu}p_\gamma + \delta_{\gamma\nu}p_\beta)\partial_\delta p_\mu. \label{eq:polaritydynfull} 
\end{align}
\begin{align}
\partial_tQ_{\alpha\beta}=&-4D_\text{r} Q_{\alpha\beta} +\frac{\tau^{0,0}}{\xi_\text{r}}(\epsilon_{\alpha\gamma}Q_{\beta\gamma}+\epsilon_{\beta\gamma}Q_{\alpha\gamma}) \nonumber\\
&+\frac{2\(\Re(\tau^{0,1})p_\nu-\Im(\tau^{0,1})\epsilon_{\mu\nu}p_\mu\)}{\xi_\text{r}}(\epsilon_{\alpha\gamma}\delta_{\beta\delta}+\epsilon_{\beta\gamma}\delta_{\alpha\delta})\(T_{\gamma\delta\nu}+\frac{1}{4}(\delta_{\gamma\delta}p_\nu+\delta_{\gamma\nu}p_\delta+\delta_{\delta\nu}p_\gamma) \) +\mathcal O(\nabla) \label{eq:nematicdynfull}
\end{align}
\end{subequations}
\end{widetext}
In the above equations, $\epsilon_{\mu\nu}$ is the totally antisymmetric tensor in 2 dimensions. We have only written out the dynamics of the nematic order tensor $Q_{\alpha\beta}$ to lowest (i.e. 0th) order in gradients since the expressions become increasingly unwieldy and higher-order terms will not be relevant for us. However, this approach can be used to determine equations for the hydrodynamic fields up to any order in gradients and Fourier modes, for arbitrary pairwise interactions with central forces.

We observe that higher moments have higher decay rates. Indeed, rotational diffusion causes the $m$-th moment to have a decay rate $m^2D_\text{r}$ (only the $m=0$ moment, the density field, does not decay, since particle number is conserved). This justifies the closure approximation of assuming that $T_{\alpha\beta\gamma}$ and all higher moments are zero, which we employ from now on.

\subsection{Hydrodynamics of non-chiral systems}
We can simplify the above equations substantially by imposing symmetry constraints on the interaction coefficients (\cref{eq:zetatau}). In particular, the pairwise force and torque acting on the particles are non-chiral,
\begin{subequations}
\begin{align}
    F(r,\varphi,\theta)&=F(r,-\varphi,-\theta), \\
    \Gamma(r,\varphi,\theta)&=-\Gamma(r,-\varphi,-\theta).
\end{align}
\end{subequations}
Similarly, by chiral symmetry of the steady state we have $g(r,\varphi,\theta)=g(r,-\varphi,-\theta)$. For example, by applying these symmetries to $\tau^{0,1}$, we obtain
\begin{align}
    \tau^{0,1}&=\frac{1}{2\pi}\int_0^\infty r\, \dd r\int_0^{2\pi}\dd\varphi\int_0^{2\pi}\dd\theta\, \Gamma(r,\varphi,\theta)e^{-i\theta}g(r,\varphi,\theta) \nonumber \\
    &=-\frac{1}{2\pi}\int_0^\infty r\, \dd r\int_0^{2\pi}\dd\varphi\int_0^{2\pi}\dd\theta\, \Gamma(r,\varphi,\theta)e^{+i\theta}g(r,\varphi,\theta) \nonumber\\
    &=0-i\tau^{0,1}_\text{I}
\end{align}
with the real coefficient
\beq
\tau^{0,1}_\text{I}=\frac{1}{2\pi}\int_0^\infty r\, \dd r\int_0^{2\pi}\dd\varphi\int_0^{2\pi}\dd\theta\, \Gamma(r,\varphi,\theta)\sin(\theta) g(r,\varphi,\theta).
\eeq

Using similar arguments, we express the other coefficients as
\begin{subequations}
\begin{align}
\zeta^{0,0}_{\alpha\beta}&=\zeta^{0,0}\delta_{\alpha\beta}, \label{eq:zeta00}\\
\zeta^{0,1}_{\alpha\beta}&=\zeta^{0,1}_\text{R}\delta_{\alpha\beta}-i\zeta^{0,1}_\text{I}\epsilon_{\alpha\beta},\\
\zeta^{1,0}_{\alpha\beta\gamma\delta} &= \zeta^{1,0}_{\text{R}1}\delta_{\alpha\gamma}\delta_{\beta\delta}+\zeta^{1,0}_{\text{R}2}\epsilon_{\alpha\gamma}\epsilon_{\beta\delta}, \\
\zeta^{1,1}_{\alpha\beta\gamma\delta} &= \zeta^{1,1}_{\text{R}1}\delta_{\alpha\gamma}\delta_{\beta\delta}+\zeta^{1,1}_{\text{R}2}\epsilon_{\alpha\gamma}\epsilon_{\beta\delta}-i\zeta^{1,1}_\text{I}(\delta_{\alpha\gamma}\epsilon_{\beta\delta}+\delta_{\beta\delta}\epsilon_{\alpha\gamma}) \\
\tau^{0,0}&=0, \\
\tau^{0,1}&=-i\tau^{0,1}_\text{I}, \\
\tau^{1,0}_{\alpha\beta}&=\tau^{1,0}\epsilon_{\alpha\beta}, \\
\tau^{1,1}_{\alpha\beta}&=\tau_\text{R}^{1,1}\epsilon_{\alpha\beta}-i\tau^{1,1}_\text{I}\delta_{\alpha\beta},\label{eq:tau11}
\end{align}
\end{subequations}
where all newly introduced coefficients are real, and the subscripts $R$ and $I$ refer to the real and imaginary parts of the original complex coefficients. 

\subsubsection{Density}
By inserting these coefficients into \cref{eq:hydrodynamic_eq_general}, contracting tensors and grouping terms together, we obtain the following equation for the density field:
\begin{align}
    \partial_t\rho=&-\partial_\alpha J_\alpha \label{eq:densdyn}, \\
    J_\alpha=&\, v_{\alpha\beta}[\rho,Q]p_\beta-D_{\alpha\beta}[\rho,Q]\partial_\beta \rho \nonumber \\&+ \kappa_1 \partial_\beta (p_\alpha p_\beta) + \kappa_2\partial_\alpha(p^2). \label{eq:denscurrent}
\end{align}
The hydrodynamic coefficients appearing in \cref{eq:denscurrent} are combinations of the different coefficients from \crefrange{eq:zeta00}{eq:tau11} and the other microscopic parameters of the system, like the self-propulsion speed $v_0$, the translational and rotational diffusivities $D_\text{t}$ and $D_\text{r}$, and the translational and rotational friction coefficients $\xi_\text{t}$ and $\xi_\text{r}$.

We observe that the density field is advected by the polarization field with an anisotropic velocity
\begin{subequations}
\begin{align}
v_{\alpha\beta}[\rho,Q]&=\(v_0-\frac{\zeta^{0,0}+\zeta_\text{R}^{0,1}+\zeta_I^{0,1}}{\xi_\text{t}}\rho\)\delta_{\alpha\beta}-\frac{2\zeta_\text{R}^{0,1}}{\xi_\text{t}}Q_{\alpha\beta} \label{eq:coeff_v_tens}\\
&=v_\text{iso}[\rho]\delta_{\alpha\beta}-v_\text{anis} Q_{\alpha\beta}
\end{align}
\end{subequations}
and diffuses according to the anisotropic diffusion tensor
\begin{subequations}
\begin{align}
D_{\alpha\beta}[\rho]&=\(D_\text{t}+\frac{\zeta_{\text{R}1}^{1,0}+\zeta_{\text{R}2}^{1,0}}{2\xi_\text{t}}\rho\)\delta_{\alpha\beta}+\frac{\zeta_{\text{R}1}^{1,0}-\zeta_{\text{R}2}^{1,0}}{2\xi_\text{t}}Q_{\alpha\beta} \label{eq:coeff_D_tens}\\
&=D_\text{iso}[\rho]\delta_{\alpha\beta}+D_\text{anis} Q_{\alpha\beta}.
\end{align}
\end{subequations}
It is well established that repulsive forces, encoded here in the coefficients $\zeta$, lead to a slowdown of active particles in dense regions, which can trigger MIPS \cite{cates2015,speck2020}. Another effect of repulsive forces is to increase effective diffusivity, since repulsion tends to homogenize the spatial distribution of particles. These effects may become anisotropic in the presence of orientational order in the system because the statistics of collisions between particles are modified by the order. For example, head-on collisions which will have a strong slow-down effect become less likely as polar order in the system increases. As a result, the advection velocity and diffusivity of the density field become anisotropic.

Two further polarization-dependent terms appear in the density current, with coefficients
\begin{subequations}
\begin{align}
\kappa_1 &=\frac{\zeta^{1,1}_{\text{R}2}-\zeta^{1,1}_{\text{R}1}-2\zeta^{1,1}_\text{I}}{2\xi_\text{t}}, \label{eq:coeff_kappa1}\\
\kappa_2 &=\frac{3\zeta^{1,1}_{\text{R}2}-\zeta^{1,1}_{\text{R}1}+2\zeta^{1,1}_\text{I}}{4\xi_\text{t}}. \label{eq:coeff_kappa2}
\end{align}
\end{subequations}
These terms describe density currents caused by gradients in polar order. 

\subsubsection{Polarization}
For a system with chiral symmetry, the polarization dynamics simplify to
\begin{align}
    \partial_t p_\alpha &+\lambda_1 p_\beta\partial_\beta p_\alpha + \lambda_2(\partial_\beta p_\beta)p_\alpha + \lambda_3\partial_\alpha(p^2) \nonumber \\ =&\,  a[\rho]p_\alpha-b|p^2|p_\alpha-\partial_\alpha P[\rho] \nonumber\\ &+\eta[\rho]\partial_\beta\partial_\beta p_\alpha +\eta_\text{B}[\rho]\partial_\alpha(\partial_\beta p_\beta) \nonumber\\
    &-\partial_\beta(w[\rho]Q_{\alpha\beta})-\nu Q_{\alpha\beta}\partial_\beta\rho-\gamma Q_{\alpha\beta} p_\beta \nonumber \\
    &+\mu_1 (\partial_\beta\partial_\beta\rho)p_\alpha+\mu_2 p_\beta\partial_\beta \partial_\alpha\rho \nonumber\\ &+\mu_3\big( (\partial_\alpha\rho)(\partial_\beta p_\beta) + (\partial_\beta\rho)(\partial_\alpha p_\beta) \big)+\mu_4(\partial_\beta\rho)(\partial_\beta p_\alpha) \nonumber \\
    &+\mu_5\partial_\gamma\big(Q_{\gamma\beta}\partial_\beta p_\alpha+\delta_{\alpha\gamma}Q_{\beta\gamma}\partial_\beta p_\gamma+Q_{\gamma\beta}\partial_\alpha p_\beta\big) \nonumber \\ &+\mu_6\partial_\gamma\big(Q_{\alpha\beta}\partial_\gamma p_\beta\big)+\mu_7\partial_\gamma\big(Q_{\alpha\gamma}\partial_\beta p_\beta+Q_{\alpha\beta}\partial_\beta p_\gamma\big) \label{eq:polaritydyn}.
\end{align}
Many of these terms have been extensively discussed in the flocking literature. The terms on the left-hand side with coefficients
\begin{subequations}
\begin{align}
\lambda_1&= -\frac{\zeta_\text{R}^{0,1}+\zeta^{0,1}_\text{I}}{\xi_\text{t}}-\frac{3\tau_\text{I}^{1,1}-\tau_\text{R}^{1,1}}{2\xi_\text{r}}, \label{eq:coeff_lambda1} \\
\lambda_2&= -\frac{\zeta_\text{R}^{0,1}+\zeta^{0,1}_\text{I}}{\xi_\text{t}}-\frac{\tau_\text{I}^{1,1}+\tau_\text{R}^{1,1}}{2\xi_\text{r}}, \label{eq:coeff_lambda2} \\
\lambda_3&= -\frac{\zeta_\text{R}^{0,1}-\zeta^{0,1}_\text{I}}{2\xi_\text{t}}-\frac{3\tau_\text{R}^{1,1}-\tau_\text{I}^{1,1}}{4\xi_\text{r}}. \label{eq:coeff_lambda3}
\end{align}
\end{subequations}
are similar to the convective non-linearities appearing in the Navier-Stokes equation. There, these coefficients are constrained to the values $\lambda_1=1$ and $\lambda_2=\lambda_3=0$ due to Galilean invariance, which is absent in this dry active matter system due to friction with the substrate \cite{toner1995,marchetti2013}.

Competition between alignment interactions and decay of alignment due to rotational diffusion leads to a density-dependent polarity growth rate
\beq \label{eq:arho}
a[\rho]=\frac{\tau_\text{I}^{0,1}}{\xi_\text{r}}\rho-D_\text{r}.
\eeq
The system undergoes the flocking transition when $a$ turns positive, i.e., when alignment arising from torques in $\tau_\text{I}^{0,1}$ overcomes rotational noise quantified by $D_\text{r}$. When $a>0$, the polarity grows exponentially. To saturate this growth, we phenomenologically introduced the non-linear term $-b|p^2|p_\alpha$, also present in the original Toner-Tu equations \cite{toner1995}, which does not emerge from our coarse-graining procedure. As such, we do not have an expression for the parameter $b$ in terms of the microscopic parameters of our system. Since we only consider pairwise interactions, terms which are cubic in the hydrodynamic fields do not appear in the final equations. This does not pose an issue, however, since this non-linear term does not affect the linear stability of the steady states that we are interested in. 

The next term is a pressure-like term,
\beq \label{eq:pressure}
P[\rho]=\frac{1}{2}\(v_0-\[ \frac{\zeta^{0,0}}{\xi_\text{t}}+\frac{\tau^{1,0}}{2\xi_\text{r}} \]\rho\)\rho,
\eeq
due to the activity of the particles. Repulsive forces that cause a density-induced slowdown can decrease this active pressure, which is described by the term proportional to $\zeta^{0,0}$. If torques cause particles to turn up (down) density gradients, pressure will be decreased (increased). This effect is encoded in the coefficient $\tau^{1,0}$ (see Ref.~\cite{zhang2021}).

Furthermore, we see density-dependent terms mirroring shear and bulk viscosities in the Navier-Stokes equation,
\begin{subequations}
\begin{align}
\eta[\rho]&=D_\text{t}+\frac{1}{4\xi_\text{t}}(\zeta^{1,1}_{\text{R}1}+3\zeta^{1,1}_{\text{R}2}-2\zeta^{1,1}_\text{I})\rho, \label{eq:coeff_eta} \\
\eta_\text{B}[\rho]&=\frac{1}{2\xi_\text{t}}(\zeta^{1,1}_{\text{R}1}-\zeta^{1,1}_{\text{R}2}+2\zeta^{1,1}_\text{I})\rho. \label{eq:coeff_etaB}
\end{align}
\end{subequations}

The other terms that appear in \cref{eq:polaritydyn} are absent from typical formulations of the Toner-Tu equations. The most important of these is the active nematic term $-\partial_\beta(w[\rho]Q_{\alpha\beta})$ with density-dependent magnitude
\beq
w[\rho]=v_0-\frac{\zeta^{0,0}}{\xi_\text{t}}\rho. \label{eq:coeff_w}
\eeq

The other terms in the third line of \cref{eq:polaritydyn} couple the nematic order field to density gradients or polarity, with coupling strengths
\begin{subequations}
\begin{align}
\nu= \tau^{1,0}/\xi_\text{r}, \label{eq:coeff_nu} \\
\gamma=2\tau^{0,1}_\text{I}/\xi_\text{r} \label{eq:coeff_gamma},
\end{align}
\end{subequations}
respectively. The standard Toner-Tu equations neglect nematic order, which is why these terms do not appear there.

The remaining terms with coefficients $\mu_i$ are of second order in both fields and gradients, and therefore do not appear in the Toner-Tu equations either. As such, they are also not of relevance for linear stability, and we shall neglect them in the following. We spell their coefficients out here only for the sake of completeness:
\begin{subequations}
\begin{align}
    \mu_1=&\,\frac{\zeta_{\text{R}1}^{1,0}+3\zeta_{\text{R}2}^{1,0}}{4\xi_\text{t}}, \\
    \mu_2=&\,\frac{\zeta_{\text{R}1}^{1,0}-\zeta_{\text{R}2}^{1,0}}{2\xi_\text{t}},\\
    \mu_3=&\,\frac{\zeta_{\text{R}1}^{1,1}-\zeta_{\text{R}2}^{1,1}+2\zeta_\text{I}^{1,1}+\zeta_{\text{R}1}^{1,0}-\zeta_{\text{R}2}^{1,0}}{4\xi_\text{t}},\\
    \mu_4=&\,\frac{\zeta_{\text{R}1}^{1,1}+3\zeta_{\text{R}2}^{1,1}-2\zeta_\text{I}^{1,1}+\zeta_{\text{R}1}^{1,0}+3\zeta_{\text{R}2}^{1,0}}{4\xi_\text{t}},\\
    \mu_5=&\,\frac{\zeta_{\text{R}1}^{1,1}-\zeta_{\text{R}2}^{1,1}-2\zeta_\text{I}^{1,1}}{3\xi_\text{t}},\\
    \mu_6=&\,\frac{\zeta_{\text{R}1}^{1,1}+5\zeta_{\text{R}2}^{1,1}-2\zeta_\text{I}^{1,1}}{3\xi_\text{t}},\\
    \mu_7=&\,\frac{\zeta_{\text{R}1}^{1,1}-\zeta_{\text{R}2}^{1,1}+4\zeta_\text{I}^{1,1}}{3\xi_\text{t}}.
\end{align}
\end{subequations}

\subsubsection{Nematic order density}
Finally, using our symmetry assumptions, the equation for the nematic order density reduces to
\begin{align} \label{eq:nematicdyn}
    \partial_t Q_{\alpha\beta}&=-4D_\text{r} Q_{\alpha\beta} + \gamma\Big(p_\alpha p_\beta-\frac{1}{2}\delta_{\alpha\beta}p^2\Big).
\end{align}
Apart from the decay term due to rotational diffusion, we see that local polarization will cause nematic order to increase with rate $\gamma=2\tau^{0,1}_\text{I}/\xi_\text{r}$. Note that this coefficient is equal to the coefficient coupling polarization and nematic order density in the polarization equation.

\section{Linear stability analysis} \label{si:lsa}
In Supplementary Note \cref{si:coarsegrain}, we derived a hydrodynamic description of the interacting active particle model in \cref{eq:eom_si}. Here, we use this theory to predict the phase diagram of our system (see \cref{fig:simulations-polarity,fig:simulations-streamparam}). To this end, we study the linear stability of the steady states of the hydrodynamic theory, and determine the instability thresholds as a function of the hydrodynamic coefficients.

\subsection{Linear stability of the uniform, isotropic state}\label{sisec:lsa_iso}

To understand how order can emerge in our system, we begin by analyzing the disordered steady state of our system. In this scenario, the density is homogeneous, $\rho(\vec r,t)=\rho_0$ and there is no polar or nematic order, $p_\alpha(\vec r,t)=0$ and $Q_{\alpha\beta}(\vec r,t)=0$. We now admit small perturbations to this solution, $\rho(\vec r,t)=\rho_0+\delta\rho(\vec r,t)$, $p_\alpha(\vec r,t)=0+\delta p_\alpha(\vec r,t)$ and $Q_{\alpha\beta}(\vec r,t)=0+\delta Q_{\alpha\beta}(\vec r,t)$. To simplify notation, we will not explicitly write out the space and time dependence of the perturbations in the following.

We now linearize \cref{eq:densdyn,eq:polaritydyn,eq:nematicdyn} around this steady state, keeping contributions only to linear order in the perturbations. From \cref{eq:nematicdyn}, we immediately obtain $\partial_t \delta Q_{\alpha\beta}=-4D_\text{r}\delta Q_{\alpha\beta}$,
implying that perturbations of the nematic order field decay to zero, $\delta Q_{\alpha\beta}\approx0$. The remaining two equations inform us about the evolution of the density and polarization perturbations,
\begin{subequations}
\begin{align}
\partial_t\delta\rho =&\,D_\text{iso}[\rho_0]\partial_\alpha\partial_\alpha\delta\rho-v_\text{iso}[\rho_0]\partial_\alpha\delta p_\alpha, \\
\partial_t \delta p_\alpha =&\,-P'[\rho_0]\partial_\alpha\delta \rho+\big(a[\rho_0]+\eta[\rho_0]\partial_\beta\partial_\beta\big)\delta p_\alpha \nonumber\\ &+\eta_{B}[\rho_0]\partial_\alpha\partial_\beta\delta p_\beta,
\end{align}
\end{subequations}
where $P'=\delta P/\delta\rho$ is the inverse compressibility.

In the long-wavelength limit, in which we can neglect all gradients, only the linear growth term $\partial_t \delta p_\alpha =a[\rho_0]\delta p_\alpha $ remains. This will trigger an instability of the polarization field if the growth rate $a[\rho_0]$ becomes positive. By looking at the expression of $a[\rho_0]$ in terms of the microscopic parameters of our system, \cref{eq:arho}, we see that this transition occurs if alignment between particles due to torques overcomes the decay of polarization due to rotational fluctuations, causing the system to transition from a disordered to a flocking phase, as indicated by the magenta phase boundary in \cref{fig:simulations-polarity,fig:simulations-streamparam}.

On the other hand, if $a[\rho_0]<0$, polarization fluctuations decay and become enslaved to the density fluctuations,
\beq
\delta p_\alpha\approx\frac{P'[\rho_0]}{a[\rho_0]}\partial_\alpha\delta\rho +\mathcal O(\nabla^2),
\eeq
which we insert into the dynamics of $\delta\rho$. This leads to the effective diffusion equation $\partial_t \delta\rho=D_\text{eff}[\rho_0]\nabla^2\delta\rho$ with
\beq \label{eq:effdiff}
D_\text{eff}[\rho_0]=D_\text{iso}[\rho_0]+\frac{v_\text{iso}[\rho_0]P'[\rho_0]}{|a[\rho_0]|}.
\eeq
A spinodal instability of the density field occurs if $D_\text{eff}[\rho_0]<0$, triggering motility-induced phase separation. By considering the expression of the inverse compressibility obtained from our coarse-graining procedure,
\beq \label{eq:invcomp}
P'[\rho]=\frac{v_0}{2}- \( \frac{\zeta^{0,0}}{\xi_\text{t}}+\frac{\tau^{1,0}}{2\xi_\text{r}} \)\rho,
\eeq
we observe that phase separation can be triggered by both repulsive forces \cite{cates2015} and torques which turn particles towards regions of higher density \cite{zhang2021}.

\begin{figure}[t!]
	\centering
	\includegraphics[width=0.45\textwidth]{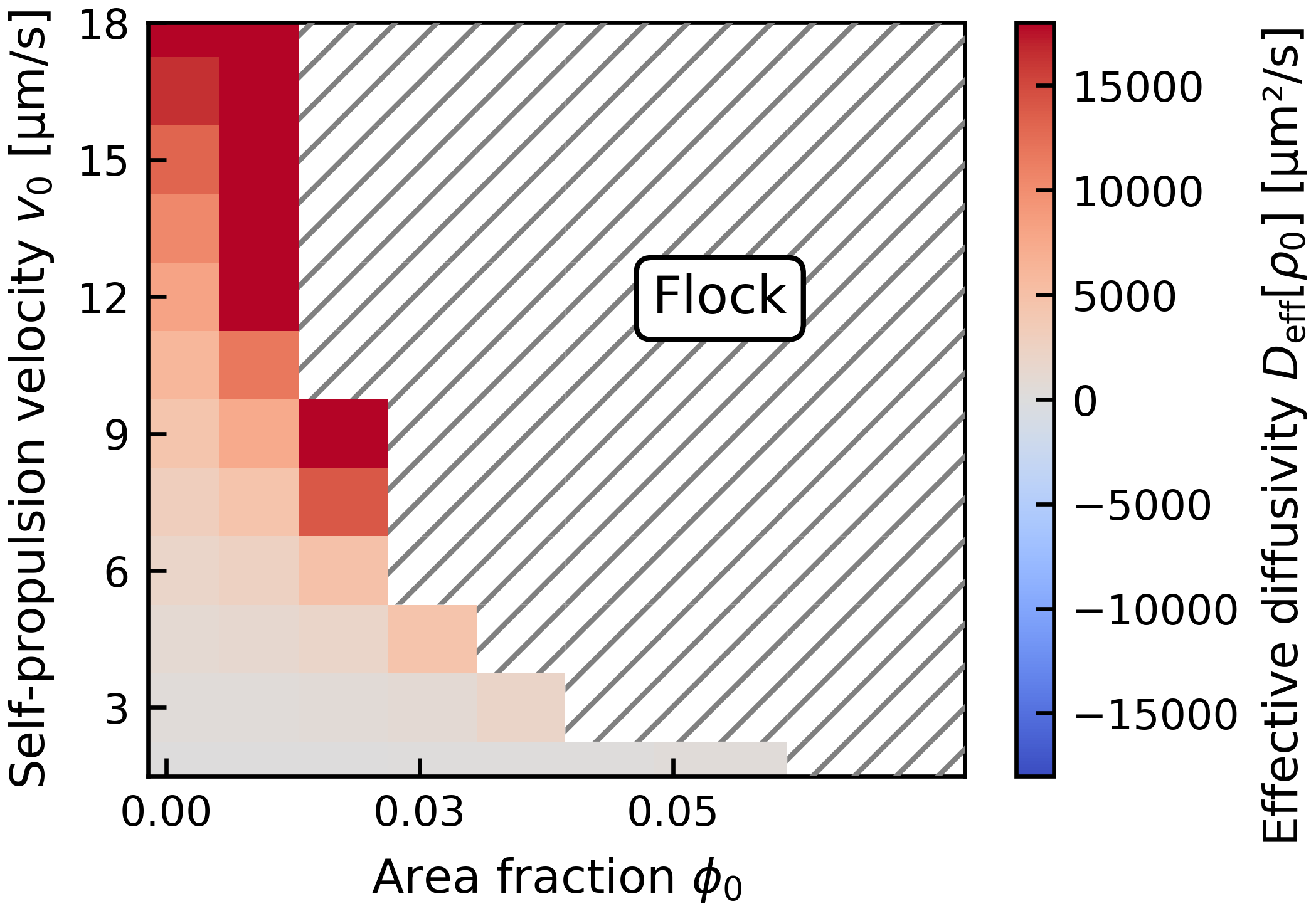}
	\caption{
		\textbf{Absence of MIPS in the isotropic state.}
		We plot the effective diffusivity $D_\text{eff}[\rho_0]$ (\cref{eq:effdiff}) in the isotropic phase (where polarity fluctuations decay, $a[\rho_0]<0$), evaluated as explained in the \hyperref[sec:methods]{Methods}. Since the diffusivity always is positive, $D_\text{eff}[\rho_0]>0$, the system does not undergo MIPS in the disordered state. The color scale is capped at $\pm1.8\cdot10^4$ $\mu$m$^2$/s for clarity.
	}  
	\label{sifig:no_mips}
\end{figure}

We obtain estimates for the polarization growth rate \cref{eq:arho} and effective diffusivity \cref{eq:effdiff} using the methods explained in the Main Text and \hyperref[sec:methods]{Methods}. We find that in our system, $a[\rho_0]$ eventually becomes positive as density and particle self-propulsion are increased, triggering a flocking transition (\cref{fig:simulations-polarity,fig:simulations-streamparam}). On the other hand, $D_\text{eff}[\rho_0]$ remains positive in the isotropic phase (\cref{sifig:no_mips}).

We deduce that the Janus colloid dimers are susceptible to start flocking, but do not phase separate in the isotropic state. This, however, does not explain the formation of streams. Streams might form as the result of a flock becoming unstable to density perturbations. Therefore, we now investigate the linear stability of the flocking state.

\subsection{Linear stability of the uniform flocking state}\label{sisec:lsa_flock}

In this section, we study the linear stability of the flocking steady state with homogeneous density $\rho(\vec r,t)=\rho_0$ and polarization density $p_\alpha(\vec r,t)=p_0e_\alpha$ along some axis $\hat{\vec x}$, such that $e_x=1$ and $e_y=0$.

As a consequence of \cref{eq:nematicdyn}, such a steady state also has a non-zero steady-state nematic-order density $Q_{\alpha\beta}^0$, given by
\beq
Q_{\alpha\beta}^0=\frac{\gamma p_0^2}{4D_\text{r}}(e_\alpha e_\beta-\frac{1}{2}\delta_{\alpha\beta})=\frac{S_0}{2}(e_\alpha e_\beta-\frac{1}{2}\delta_{\alpha\beta}),
\eeq
where we have introduced the scalar nematic order density $S_0=\gamma p_0^2/4D_\text{r}$. On the other hand, the steady state equation for the polarization density (see \cref{eq:polaritydyn}) is
\beq
0= \big((a[\rho_0]-b|p^2|)\delta_{\alpha\beta}-\gamma Q^0_{\alpha\beta}) p_\beta
\eeq
from which we deduce $p_0^2=a[\rho_0]/(b+\gamma^2/8D_\text{r})$.

As in \cref{sisec:lsa_iso}, we introduce small perturbations $\rho(\vec r,t)=\rho_0+\delta\rho(\vec r,t)$, $p_\alpha(\vec r,t)=p_0e_\alpha+\delta p_\alpha(\vec r,t)$ and $Q_{\alpha\beta}(\vec r,t)=Q_{\alpha\beta}^0+\delta Q_{\alpha\beta}(\vec r,t)$, and linearize the hydrodynamic equations. We only keep terms up to linear order in gradients, since this is sufficient to identify the instability of interest:
\begin{subequations}
\begin{align}
    \partial_t&\delta\rho=  -v_{\alpha\beta}\partial_\alpha\delta p_\beta-p_0 v_{xx}'\partial_x\delta\rho +p_0v_\text{anis}\partial_\alpha\delta Q_{\alpha x}, \label{eq:deltarho}\\
    \partial_t&\delta p_x= \nonumber\\& a'p_0\delta\rho-\(2a-\frac{\gamma^2p_0^2}{8D_\text{r}}\) \delta p_x-\(P'+\frac{w'+\nu}{2}S_0\)\partial_x\delta\rho \nonumber\\& -w\partial_\alpha\delta Q_{\alpha x} -\gamma p_0 \delta Q_{xx} \nonumber\\&-(\lambda_1+2\lambda_3)p_0 \partial_x\delta p_x-\lambda_2p_0\partial_\alpha \delta p_\alpha,\\
    \partial_t&\delta p_y= \nonumber\\& -\(P'-\frac{w'+\nu}{2}S_0\)\partial_y\delta\rho+\gamma (S_0\delta p_y - p_0 \delta Q_{xy}) \nonumber\\& -w\partial_\alpha\delta Q_{\alpha y}-\lambda_1p_0 \partial_x\delta p_y-2\lambda_3p_0\partial_y \delta p_x, \label{eq:deltapy}\\
    \partial_t&\delta Q_{\alpha\beta}= \nonumber\\&-4D_\text{r}\delta Q_{\alpha\beta}+\gamma p_0(e_\alpha\delta p_\beta+e_\beta \delta p_\alpha -\delta_{\alpha\beta}\delta p_x)
\end{align}
\end{subequations}
If not otherwise specified, all functionals are evaluated at steady-state conditions, i.e. $v_{\alpha\beta}=v_{\alpha\beta}[\rho_0,S_0]$ etc. Functional derivatives with respect to the density field are denoted with a prime, $v'_{\alpha\beta}=\delta v_{\alpha\beta}/\delta\rho$.

We now observe that the polarization fluctuations $\delta p_x$, which change the polarization magnitude, and nematic order fluctuations $\delta Q_{\alpha\beta}$ have a finite decay rate even in the infinite-wavelength limit in which all gradients can be neglected. These fluctuations decay quickly compared to the other, slowly varying fluctuations $\delta\rho$ and $\delta p_y$, and become enslaved to them. To lowest order in gradients, we obtain
\begin{subequations}
\begin{align}
    \delta p_x\approx&\,\frac{a'p_0}{2a-\gamma^2p_0^2/4D_\text{r}}\delta\rho=\alpha\delta\rho,\\
    \delta Q_{\alpha\beta}\approx&\,\frac{\gamma p_0}{4D_\text{r}}(e_\alpha \delta p_\beta +e_\beta\delta p_\alpha-\delta_{\alpha\beta}\delta p_x).
\end{align}
\end{subequations}

We can now insert these expressions into \cref{eq:deltarho,eq:deltapy}. Note that this causes the term $S_0\delta p_y - p_0 \delta Q_{xy}$ in \cref{eq:deltapy} to cancel, proving that $\delta p_y$ is indeed a slow mode. We can then recast  the equations for the slow modes into the form
\begin{subequations}
\begin{align}
    \partial_t\delta\rho=&\, -m_{\rho\rho}\partial_x\delta\rho -A\partial_y\delta p_y, \label{eq:deltarho_linearized}\\
    \partial_t\delta p_y=&\, -B\partial_y\delta\rho-m_{pp}\partial_x\delta p_y. \label{eq:deltapy_linearized}
\end{align}
\end{subequations}
with
\begin{subequations}
\begin{align}
    m_{\rho\rho}&=p_0v_{xx}'+\alpha\(v_{xx}-\frac{\gamma v_\text{anis} p_0^2}{4D_\text{r}}\), \\
    A&= v_{yy}-\frac{\gamma v_\text{anis} p_0^2}{2D_\text{r}}, \label{eq:A}\\
    B&= P'-\frac{\gamma (w'+\nu)p_0^2}{8D_\text{r}}+\alpha p_0\(2\lambda_3 -\frac{\gamma w}{4D_\text{r}}\), \label{eq:B_si}\\
    m_{pp}&= \(\lambda_1+\frac{\gamma w}{4D_\text{r}}\)p_0
\end{align}
\end{subequations}
In Fourier space, these coupled equations read
\beq
\partial_t\begin{pmatrix}
	\delta\rho\\
	\delta p_\perp\\
\end{pmatrix}= -ik\begin{pmatrix}
	m_{\rho\rho}\cos\psi & A\sin \psi \\
	B\sin\psi & m_{p p}\cos \psi\\
\end{pmatrix}
\cdot\begin{pmatrix}
	\delta\rho\\
	\delta p_\perp\\
\end{pmatrix}
\eeq
where $\psi$ is the angle between the wavevector $\vec k$ and the $x$ axis, $\vec k=k(\cos\psi,\sin\psi)^T$.

The stability of the flock to fluctuations is determined by the eigenvalues of the stability matrix. If an eigenvalue acquires a positive real part, the system is linearly unstable. In our case, the eigenvalues are
\begin{align}
\lambda_\pm=&\,-\frac{ik}{2}(m_{\rho\rho}+m_{pp})\cos\psi \nonumber \\&\,\pm \frac{ik}{2}\sqrt{(m_{\rho\rho}-m_{pp})^2\cos^2\psi+4AB\sin^2\psi}. \label{eq:evals_stream_instab}
\end{align}
From this, we see that an instability occurs if the discriminant becomes negative, which is the case if
\beq
B<-\frac{(m_{\rho\rho}-m_{pp})^2}{4A\tan^2\psi}.
\eeq
The most unstable wavevector is the one perpendicular to the direction $\hat{\vec x}$ of flocking, with $\psi=\pi/2$. It is unstable if $B<0$. The quantity $B$ acts as an effective inverse compressibility of the flock \cite{miller2024}. If it becomes negative, high-density regions will effectively have a lower pressure than low-density regions, causing a further inflow of particles into denser regions, triggering an instability.

A more thorough analysis of this instability, in the simpler case where nematic order is assumed to be negligible ($\gamma=0$), can be found in \cite{miller2024,miller2024a}. Our results show, however, that nematic order produces an additional instability mechanism, which we call active nematic condensation. As discussed in the Main Text, this mechanism is the main driver of the stream formation in our system; therefore, nematic order can not be neglected in our analysis.

\section{Supplementary Movies}

\textbf{Movie 1: Stream of Janus particle dimers.}\label{mov:expt_stream} Stream of Janus particle dimers imaged with a $10\times$ objective at 17 frames per second. The particle area fraction is $\phi_0 = 0.22\pm 0.01$ and the self-propulsion speed is $v_0= 8.44\pm0.02$ $\mu$m/s.

\bigskip

\textbf{Movie 2: Stream of Janus dimers flowing into a high-density flock.}\label{mov:expt_vortex} Stream of Janus particle dimers flowing into a high-density flock with large-scale vortices, imaged with a $10\times$ objective at 17 frames per second.

\bigskip

\textbf{Movie 3: Isotropic state in simulations.}\label{mov:sim_iso} Simulation showing the isotropic state, shown in \cref{fig:simulations-isotropic}. The simulation contains $N=10000$ dimers with global area fraction $\phi_0=0.02$ and self-propulsion speed $v_0=3$ $\mu$m/s. 

\bigskip

\textbf{Movie 4: Traveling bands in simulations.}\label{mov:sim_band} Simulation showing traveling bands, shown in \cref{fig:simulations-bands}. The simulation contains $N=10000$ dimers with global area fraction $\phi_0=0.02$ and self-propulsion speed $v_0=15$ $\mu$m/s.

\bigskip

\textbf{Movie 5: Uniform flock in simulations.}\label{mov:sim_flock} Simulation showing the uniform flock state, shown in \cref{fig:simulations-flock}. The simulation contains $N=10000$ dimers with global area fraction $\phi_0=0.07$ and self-propulsion speed $v_0=15$ $\mu$m/s.

\bigskip

\textbf{Movie 6: Streams surrounded by flocking phase in simulations.}\label{mov:sim_stream_flock} Simulation showing streams surrounded by a dilute flocking phase. The simulation contains $N=10000$ dimers with global area fraction $\phi_0=0.12$ and self-propulsion speed $v_0=15$ $\mu$m/s.

\bigskip

\textbf{Movie 7: Streams surrounded by isotropic phase in simulations.}\label{mov:sim_stream_iso} Simulation showing streams surrounded by an isotropic phase, shown in \cref{fig:simulations-stream}. The simulation contains $N=10000$ dimers with global area fraction $\phi_0=0.175$ and self-propulsion speed $v_0=15$ $\mu$m/s.

\end{document}